\documentclass[aps,prd,superscriptaddress,twoside,twocolumn,nofootinbib,showpacs,
floatfix]{revtex4-2}
\usepackage{amsmath,amssymb}
\usepackage{graphicx,bm}
\usepackage{slashed,color}
\usepackage[colorlinks=true,pdfstartview=FitV,bookmarks=true,bookmarksnumbered=true,
breaklinks]{hyperref}

\hypersetup{linkcolor=red, citecolor=blue, urlcolor=blue}

\usepackage[normalem]{ulem} 
\usepackage[dvipsnames]{xcolor} 
\begin{document}
\title{Kaon-induced $\phi \Sigma$ production off the proton}
\author{Sang-Ho Kim}
\email{shkimphy@gmail.com}
\affiliation{Department of Physics and Origin of Matter and Evolution of Galaxies
(OMEG) Institute, Soongsil University, Seoul 06978, South Korea}
\author{Myung-Ki Cheoun}
\affiliation{Department of Physics and Origin of Matter and Evolution of Galaxies
(OMEG) Institute, Soongsil University, Seoul 06978, South Korea}

\date{\today}

\begin{abstract}

We investigate the reaction mechanism of $K^-p \to \phi\Sigma^0$ within a hybrid
Regge approach based on effective Lagrangians.
The nonresonant background includes Reggeized $t$-channel $K$ and $K^*$ exchanges,
together with ground-state $\Sigma$ and nucleon exchanges in the $s$ and $u$ channels,
respectively.
To describe the structures observed at $3.0 \leqslant P_{\rm Lab} \leqslant 4.5$ GeV,
we include the high-mass $\Sigma(2620)$ and $\Sigma(3000)$ resonances and examine the
spin-parity assignments $J^P = 1/2^\pm$, $3/2^\pm$, and $5/2^\pm$.
The $K^*$-Reggeon exchange dominates the forward-angle cross sections, whereas the
smaller $K$-Reggeon contribution is important for the total cross section at low
energies and for the spin-density matrix elements.
The nonresonant background alone is not sufficient to reproduce the local structures
in the total cross section or the differential cross sections at large $-t'$.
Including the two resonances, particularly the $\Sigma(3000)$, substantially improves
the agreement with the available data.
The $J^P=5/2^-$ assignment for the $\Sigma(3000)$ provides a reasonable overall
description, although the present data do not permit a definitive determination.

\end{abstract}

\maketitle
\section{Introduction}

Since the $\phi$ meson is predominantly composed of an $s\bar{s}$ pair, its
production is expected to be sensitive to hidden-strangeness dynamics and to
mechanisms beyond ordinary nonstrange-hadron production.
The Okubo-Zweig-Iizuka (OZI) rule~\cite{Okubo:1963fa,Okubo:1963fa2,Okubo:1963fa3}
implies that the coupling of the $\phi$ meson to the nucleon is weak, owing to the
small $s\bar{s}$ component in the nucleon wave function~\cite{Henley:1991ge,
Titov:1997qz,Titov:1998bw,Armstrong:2012bi}.
Consequently, $\phi$-meson production off the nucleon is expected to exhibit features
distinct from those observed in the production of $\rho$ and $\omega$ mesons.

In $\rho$~\cite{Ballam:1971yd,CLAS:2001zxv,Wu:2005wf,GlueX:2023fcq,Wang:2025rvr} and
$\omega$~\cite{Oh:2000zi,Oh:2002rb,Titov:2002iv,Sibirtsev:2003qh,Denisenko:2016ugz,
Yu:2018ydp,Wei:2019imo} photoproduction, $s$-channel $N^*$ resonances can provide
sizable contributions to the cross sections, in addition to the Pomeron-exchange
contribution that dominates at high energies.
By contrast, the effects of $N^*$ resonances are expected to be suppressed in
$\phi$ photoproduction~\cite{Kim:2019kef,Kim:2021adl}.
Nevertheless, several studies have attempted to incorporate $N^*$
resonances~\cite{Kiswandhi:2010ub,Kiswandhi:2011cq,Wu:2023ywu} and rescattering
effects~\cite{Ozaki:2009mj,Ryu:2012tw} to account for the bump structure around
$W \simeq 2.2\, \mathrm{GeV}$ at forward angles observed by the
LEPS~\cite{LEPS:2005hax} and CLAS~\cite{CLAS:2013jlg,Dey:2014tfa} Collaborations.

$\phi$-meson production induced by pion beams is also of considerable interest.
The J-PARC P95 proposal was submitted to study the $\pi^- p \to \phi n$
reaction~\cite{J-PARC:P95}, with the primary aim of investigating whether a bump
structure similar to that observed in $\gamma p \to \phi p$ can also appear in this
hadronic reaction.
Previous studies have shown that the inclusion of $s$-channel $N^*$ resonances can
be important for understanding the energy dependence of the $\pi^- p \to \phi n$
cross section~\cite{Chung:1997tu,Doring:2008sv,Xie:2007qt,Wang:2024qnk,Lee:2026kcd},
suggesting that this reaction can provide useful information on $N^*$ resonances
that couple to the $\phi N$ channel.

It is then natural to extend this framework to the strangeness sector through
kaon-induced reactions~\cite{Takahashi:2012cka,Agari:2012gj}, such as the $K^- p \to
\phi \Lambda$ and $K^- p \to \phi \Sigma^0$ processes, where intermediate hyperon
resonances can contribute through the $s$ channel.
The $\phi \Lambda$ and $\phi \Sigma$ final states serve as isospin filters for
$\Lambda^*$ and $\Sigma^*$ resonances, respectively.
One of the authors previously studied the $K^- p \to \phi \Lambda$ reaction within a
Regge-based effective Lagrangian approach~\cite{Kim:2026tgd}.
It was found that the total and differential cross sections, as well as the
spin-density matrix elements (SDMEs), are mainly described by the $t$-channel $K$-
and $K^*$-Reggeon exchanges, while the contributions from the $s$-channel $\Lambda^*$
resonances are negligible.
Whether the same conclusion holds for the isospin-partner reaction $K^- p \to \phi
\Sigma^0$ is therefore worth investigating.

In this work, we study the reaction mechanism of the $K^- p \to \phi \Sigma^0$
reaction within a similar framework.
It is interesting to note that the total cross section for $K^- p \to \phi
\Sigma^0$~\cite{Lindsey:1966zz,Lyons:1977rp,Aguilar-Benitez:1972ngz,Rouge:1972xr,
Ayres:1974aj,ACNO:1977ctz,DeGroot:1974qw} displays a markedly different energy
dependence from that for $K^- p \to \phi \Lambda$: pronounced local structures appear
in the incident $K^-$ laboratory (Lab) momentum range $3.0 \leqslant P_{\mathrm{Lab}}
\leqslant 4.5\,\mathrm{GeV}$, corresponding to the center-of-mass (c.m.) energy range
$2.6 \leqslant W \leqslant 3.1\,\mathrm{GeV}$.
This feature suggests that additional reaction mechanisms beyond the background $K$-
and $K^*$-Reggeon exchanges may be required in the $\phi \Sigma$ channel.

\begin{figure*}[ht]
\centering
\includegraphics[scale=0.55]{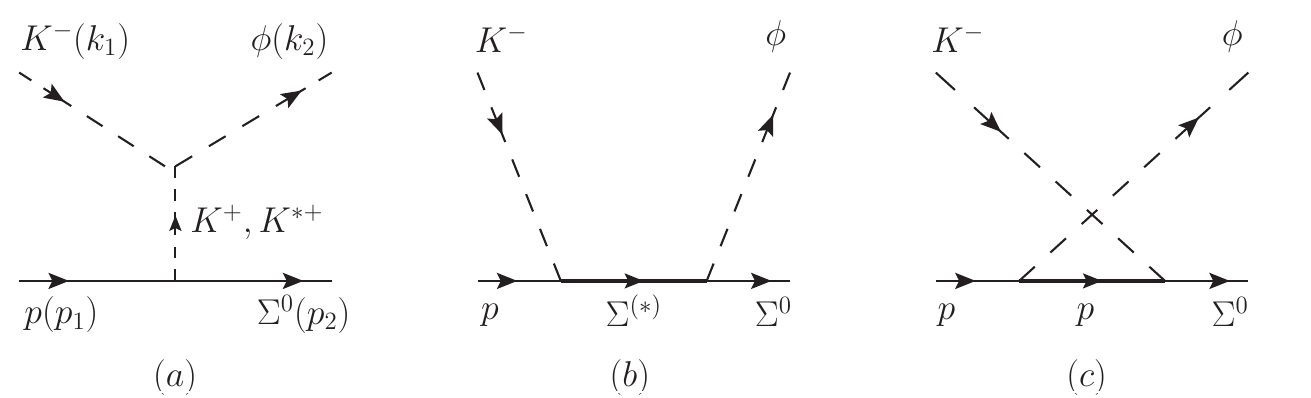}
\caption{Feynman diagrams for the $K^- p \to \phi \Sigma^0$ reaction:
(a) $t$-channel $K$- and $K^*$-Reggeon exchanges, (b) $s$-channel ground-state
$\Sigma$ and $\Sigma^*$ resonance exchanges, and (c) $u$-channel nucleon exchange.}
\label{FIG01}
\end{figure*}
One possibility is the contribution from high-mass $\Sigma^*$ resonances.
The Particle Data Group (PDG)~\cite{PDG:2024cfk} lists a few relevant candidates,
including $\Sigma(2620)$ and $\Sigma(3000)$, although their spin-parity quantum
numbers are not yet firmly established.
A coupled-channel study predicts only relatively low-mass $\Sigma^*$ resonances
that couple strongly to the $\phi \Sigma$ channel, such as $\Sigma(1670)3/2^-$ and
$\Sigma(1900)1/2^-$~\cite{Khemchandani:2018amu}.
Hidden-gauge vector-baryon calculations likewise predict relatively low-mass
$I=1$ $\Sigma^*$-like states with non-negligible couplings to the $\phi \Sigma$
channel~\cite{Oset:2010tof,Garzon:2012np}.
Another plausible mechanism is a rescattering process involving high-mass
meson-baryon intermediate channels, such as $K\Xi(2030)$, $K_1(1270)\Xi$,
$\phi\Sigma(1670)$, $K^*\Xi(1820)$, and $\phi\Sigma(1775)$.
To our knowledge, no dedicated theoretical study has yet addressed the role of such
rescattering contributions in the $\phi\Sigma$ production process, and this
mechanism remains largely unexplored.

In the present work, we focus on the former possibility and investigate
the $s$-channel $\Sigma(2620)$ and $\Sigma(3000)$ contributions in detail.
We perform a simultaneous analysis of the total and differential cross sections,
together with the SDMEs, in order to examine possible spin-parity assignments of
these states.
The results of this study may help to shed light on the spectroscopy of high-mass
hyperon resonances.

This paper is organized as follows.
In Sec.~\ref{Sec:II}, we present the theoretical framework, including the effective
Lagrangians, Reggeized $t$-channel exchanges, and $s$-channel $\Sigma^*$-resonance
amplitudes.
In Sec.~\ref{Sec:III}, we show the numerical results and discuss the role of the
high-mass $\Sigma^*$ resonances in some detail.
Finally, a summary and conclusions are given in Sec.~\ref{Sec:IV}.

\section{Theoretical Framework}
\label{Sec:II}

In this section, we briefly describe the theoretical framework employed in this work.
We introduce a hybrid Regge model that combines the effective Lagrangian approach
with Regge phenomenology.
The Feynman diagrams for the $K^- p \to \phi \Sigma^0$ reaction are shown in
Fig.~\ref{FIG01}, where the four-momenta of the initial $K^-$ and proton are
denoted by $k_1$ and $p_1$, respectively, while those of the final $\phi$ and
$\Sigma^0$ are denoted by $k_2$ and $p_2$.
We consider pseudoscalar $K$- and vector $K^*$-Reggeon exchanges in the $t$ channel
[Fig.~\ref{FIG01}(a)].
The ground-state $\Sigma$ and its resonances are included in the $s$ channel
[Fig.~\ref{FIG01}(b)].
The $u$-channel contribution is described by the nucleon exchange
[Fig.~\ref{FIG01}(c)].

\subsection{Born term}
\label{Sec:II-1}

We begin with the $t$-channel process of Fig.~\ref{FIG01}(a).
The effective Lagrangians for the interaction vertices involving the $\phi$ meson
are written as
\begin{align}
\mathcal L_{\phi K K} &=
-ig_{\phi K K} ( K^- \partial_\mu K^+ - \partial_\mu K^- K^+ ) \phi^\mu,
\cr
\mathcal L_{\phi K^* K} &=
{g_{\phi K^* K}} \varepsilon^{\mu\nu\alpha\beta} \partial_\mu \phi_\nu
(\partial_\alpha K^{*-}_\beta K^+ + \partial_\alpha K^{*+}_\beta K^-).
\cr
\label{eq:Lag1}
\end{align}
Here, $K$, $K^*$, and $\phi$ denote the fields of the $K(494,0^-)$, $K^*(892,1^-)$,
and $\phi(1020,1^-)$ mesons, respectively.
The coupling constant $g_{\phi K K} = 4.48$ is determined from the branching ratio
${\mathcal B} (\phi \to K^+ K^-) = 49.9\,\%$, together with the total width
$\Gamma_\phi =$ 4.249 MeV~\cite{PDG:2024cfk}.
The coupling $g_{\phi K^* K}$ is obtained from the SU(3) flavor-symmetry relation
\begin{align}
g_{\phi K^* K} = g_{\rho \omega \pi} / \sqrt2,
\end{align}
where $g_{\rho \omega \pi}$ is evaluated within the hidden gauge approach as
\begin{align}
g_{\rho \omega \pi} = \frac{N_c g_{\rho \pi \pi}^2}{8 \pi^2 f_\pi} = 14.4 \,
\mathrm{GeV}^{-1},
\end{align}
with $N_c = 3$, $f_\pi = 93$ MeV, and $g_{\rho \pi \pi} = 5.94$~\cite{Oh:2004wp}.
The value of $g_{\rho \pi \pi}$ is determined from $\mathcal{B} (\rho \to \pi \pi)
\sim$ 1~\cite{PDG:2024cfk}.

The effective Lagrangians for the $K^{(*)} N \Sigma$ interaction vertices are written
as
\begin{align}
\mathcal L_{K N \Sigma} &=
\frac{g_{K N \Sigma}}{M_N+M_\Sigma} \bar N \gamma_\mu
\gamma_5\, \bm{\tau} \cdot \bm{\Sigma}\, \partial^\mu K + {\rm H.c.},
\cr
\mathcal L_{K^* N \Sigma} &=
-g_{K^* N \Sigma} \bar N \biggl[ \gamma_\mu\, \bm{\tau} \cdot \bm{\Sigma}\, -
\cr &\,
\frac{\kappa_{K^* N \Sigma}}{M_N+M_\Sigma} \sigma_{\mu\nu}\, \bm{\tau} \cdot
\bm{\Sigma}\, \partial^\nu \biggr]
K^{*\mu} + \mathrm{H.c.},
\label{eq:Lag2}
\end{align}
where $N$ and $\Sigma$ stand for the nucleon and $\Sigma(1193)$ baryon fields,
respectively.
The coupling constants are adopted from the Nijmegen soft-core model (NSC97f)~\cite{
Rijken:1998yy,Stoks:1999bz}:
\begin{align}
g_{K N \Sigma} &= 5.31,
\cr
g_{K^* N \Sigma} &= -3.52, \,\,\, \kappa_{K^* N \Sigma} = -1.14.
\label{eq:Coupl1}
\end{align}

In the $t$ channel, the Regge amplitudes are constructed using a hybrid approach,
in which the Feynman propagators for the pseudoscalar $K$- and vector $K^*$-meson
exchanges are replaced by the corresponding Regge propagators associated with their
Regge trajectories~\cite{Kim:2026tgd,Kim:2026ice}:
\begin{align}
T_K  (s,t) &= \mathcal M_K(s,t)
\left( \frac{s}{s_K} \right)^{\alpha_K(t)}
\cr
& \times \Gamma (-\alpha_K(t)) \alpha_K' F_K(t),
\cr
T_{K^*} (s,t) &= \mathcal M_{K^*}(s,t)
\left( \frac{s}{s_{K^*}} \right)^{\alpha_{K^*}(t)-1}
\cr
& \times \Gamma (1-\alpha_{K^*}(t)) \alpha_{K^*}' F_{K^*}(t),
\label{eq:RegAmpl1}
\end{align}
where the amplitudes $\mathcal M_K$ and $\mathcal M_{K^*}$ are obtained from the
effective Lagrangians in Eqs.~(\ref{eq:Lag1}) and (\ref{eq:Lag2}) as follows:
\begin{align}
\mathcal M_{K}^\mu &= 2 i
\frac{g_{\phi K K} g_{K N \Sigma}}{M_N + M_\Sigma}
\gamma_\nu \gamma_5 k_1^\mu (k_2 - k_1)^\nu,
\cr
\mathcal M_{K^*}^\mu &=
g_{\phi K^* K} g_{K^* N \Sigma} \epsilon^{\mu\nu\alpha\beta}
\cr & \times
\left [ \gamma_\nu - \frac{i\kappa_{K^* N \Sigma}}{M_N + M_\Sigma}
\sigma_{\nu \lambda}(k_2-k_1)^\lambda \right ] k_{2 \alpha} k_{1 \beta},
\label{eq:Ampl1}
\end{align}
with
\begin{align}
\mathcal M = \varepsilon_\mu^* \bar{u}_\Sigma {\mathcal M}^\mu u_N.
\label{eq:InvAmpl}
\end{align}
Here $u_N$ and $u_\Sigma$ denote the Dirac spinors of the initial nucleon and
the final $\Sigma$ baryon, respectively, and are normalized as $\bar u_B u_B = 1$.
The four-vector $\varepsilon_\mu$ represents the polarization vector of the
outgoing $\phi$ meson.

The Regge trajectories in Eq.~(\ref{eq:RegAmpl1}) are determined to be~\cite{
Guidal:1997hy}
\begin{align}
\alpha_K (t) &= -0.17 + 0.7 t,
\cr
\alpha_{K^*} (t) &= 0.34 + 0.83 t.
\label{eq:ReggeParameters}
\end{align}
The energy-scale parameters are given by $s_K = s_{K^*} = 1\,\mathrm{GeV}^2$.
The form factor is introduced to account for the finite-size effects at the
interaction vertices.
We adopt the following form:
\begin{align}
F_{K(K^*)}(t) =
\left( \frac{n\Lambda_{K(K^*)}^2}{n\Lambda_{K(K^*)}^2 - t} \right)^n,
\label{eq:FormFac_t}
\end{align}
where the cutoff masses are chosen as $\Lambda_K$ = 0.6 GeV and $\Lambda_{K^*}$ =
2.4 GeV with $n$ = 1.

The $t$-dependent differential cross section is expressed as
\begin{align}
\frac{d\sigma}{dt} = \frac{M_N M_\Sigma}{16 \pi (p_{\rm c.m.})^2 s}
\frac{1}{2}\sum_{\lambda_V,s_f,s_i}|T|^2,
\label{eq:Def:dsdt}
\end{align}
where $p_{\rm c.m.}$ denotes the magnitude of the initial kaon momentum in the c.m.
frame.
The indices $s_i$, $s_f$, and $\lambda_V$ label the spin or helicity states of
the initial nucleon, final $\Sigma$ baryon, and outgoing $\phi$ meson,
respectively.

In addition to the $t$-channel Reggeon exchanges, we also examine the contributions
from the $s$-channel $\Sigma$ exchange [Fig.~\ref{FIG01}(b)] and the $u$-channel
nucleon exchange [Fig.~\ref{FIG01}(c)].
The effective Lagrangians for the $\phi$-meson--baryon interaction vertices are
written as
\begin{align}
\mathcal{L}_{\phi \Sigma \Sigma} &= - g_{\phi \Sigma \Sigma} \bar \Sigma
\left[ \gamma_\mu \Sigma - \frac{\kappa_{\phi \Sigma \Sigma}}{2M_N} \sigma_{\mu\nu}
\Sigma \partial^\nu \right] \phi^\mu,
\cr
\mathcal{L}_{\phi N N} &= - g_{\phi N N} \bar N
\left[ \gamma_\mu N - \frac{\kappa_{\phi N N}}{2M_N} \sigma_{\mu\nu}
N \partial^\nu \right] \phi^\mu,
\label{eq:Lag3}
\end{align}
where the coupling constants are taken from the Nijmegen soft-core model
(NSC97f)~\cite{Rijken:1998yy,Stoks:1999bz}:
\begin{align}
g_{\phi \Sigma \Sigma} &= -4.78, \,\,\, \kappa_{\phi \Sigma \Sigma} = -1.43,
\cr
g_{\phi N N} &= -1.47, \,\,\, \kappa_{\phi N N} = -0.654.
\label{eq:Coupl2}
\end{align}
The invariant amplitudes corresponding to the $s$-channel $\Sigma$ and $u$-channel
nucleon exchanges are, respectively, obtained as
\begin{align}
\mathcal M_\Sigma^\mu &=
i \frac{g_{\phi \Sigma \Sigma}}{s-M_\Sigma^2} \frac{g_{K N \Sigma}}{M_N + M_\Sigma}
\left [ \gamma^\mu - \frac{i\kappa_{\phi \Sigma \Sigma}}{2 M_N} \sigma^{\mu\nu}
k_{2 \nu}
\right ]
\cr & \times
(\rlap{/}{k_1}+\rlap{/}{p_1}+M_\Sigma) \gamma^\alpha \gamma_5 k_{1\alpha},
\cr
\mathcal M_N^\mu &=
i \frac{g_{\phi N N}}{u-M_N^2}
\frac{g_{K N \Sigma}}{M_N + M_\Sigma} \gamma^\alpha \gamma_5
(\rlap{/}{p_2}-\rlap{/}{k_1}+M_N)
\cr & \times
\left [ \gamma^\mu - \frac{i\kappa_{\phi N N}}{2 M_N} \sigma^{\mu\nu} k_{2 \nu}
\right ] k_{1\alpha},
\label{eq:Ampl2}
\end{align}
with $\mathcal M = \varepsilon_\mu^* \bar{u}_\Sigma {\mathcal M}^\mu u_N$.

We introduce the following form factors for the $s$- and $u$-channel diagrams for
each vertex:
\begin{align}
F_\Sigma (s) &= \frac{\Lambda_s^4}{\Lambda_s^4 + (s - M_\Sigma^2)^2},
\cr
F_N (u) &= \frac{\Lambda_u^4}{\Lambda_u^4 + (u - M_N^2)^2},
\label{eq:FormFac_su}
\end{align}
where the cutoff masses are chosen as $\Lambda_s$ = $\Lambda_u$ = 0.8 GeV.

\subsection{Resonance term}
\label{Sec:II-2}

We include a minimal set of high-mass $\Sigma^*$ resonances, namely, $\Sigma(2620)$
and $\Sigma(3000)$, in the $s$-channel amplitude [Fig.~\ref{FIG01}(b)].
These are the only candidates listed in the PDG~\cite{PDG:2024cfk} whose masses lie
in the region $2.6 \leqslant W \leqslant 3.1$ GeV, where the local structures in the
total cross section are observed.
Since their spin-parity quantum numbers remain uncertain, we test the possible
assignments $J^P$ = $1/2^\pm$, $3/2^\pm$, and $5/2^\pm$.

The effective Lagrangians for the $K N \Sigma^*$ vertex are given
by~\cite{Kim:2024mqx}
\begin{align}
\mathcal L_{K N \Sigma^*}^{1/2^\pm} &=
-i g_{K N \Sigma^*} \bar N \Gamma^{(\pm)} \Sigma^* K + \mathrm{H.c.},
\cr
\mathcal L_{K N \Sigma^*}^{3/2^\pm} &=
\frac{g_{K N \Sigma^*}}{M_K} \bar N \Gamma^{(\mp)} \Sigma^{*\mu}
\partial_\mu K + \mathrm{H.c.},
\cr
\mathcal L_{K N \Sigma^*}^{5/2^\pm} &=
i \frac{g_{K N \Sigma^*}}{M_K^2} \bar N \Gamma^{(\pm)} \Sigma^{*\mu\nu}
\partial_\mu \partial_\nu K + \mathrm{H.c.},
\label{eq:ResLag1}
\end{align}
The off-shell terms of the Rarita-Schwinger fields are neglected because the
resonances are considered near their mass shell.
The effective Lagrangians for the $\phi \Sigma \Sigma^*$ vertex can be expressed
as~\cite{Kim:2024mqx}
\begin{align}
\mathcal{L}_{\phi \Sigma \Sigma^*}^{1/2^\pm}
&= - g_{\phi \Sigma \Sigma^*} \bar \Sigma \Gamma_\mu^{(\mp)} \Sigma \phi^\mu +
\mathrm{H.c.},
\cr
\mathcal{L}_{\phi \Sigma \Sigma^*}^{3/2^\pm}
&= -i\frac{g_{\phi \Sigma \Sigma^*}}{2M_N}
\bar \Sigma \Gamma_\nu^{(\pm)} \Sigma_\mu \phi^{\mu\nu} + \mathrm{H.c.},
\cr
\mathcal{L}_{\phi \Sigma \Sigma^*}^{5/2^\pm}
&= \frac{g_{\phi \Sigma \Sigma^*}}{(2M_N)^2}
\bar \Sigma \Gamma_\nu^{(\mp)} \Sigma_{\mu\alpha}
\partial^\alpha \phi^{\mu\nu} + \mathrm{H.c.},
\label{eq:ResLag2}
\end{align}
where $\phi^{\mu\nu} = \partial^\mu \phi^{\nu} - \partial^\nu \phi^{\mu}$ and
$\Sigma^*$ denotes the $\Sigma^*$ resonance field.
Since the $\Sigma^*$ resonance states are expected to contribute mainly near their
resonance regions, additional terms are neglected.
We use the following notation:
\begin{align}
\Gamma^{(\pm)} = \left(
\begin{array}{c}
\gamma_5 \\ \mathbf{1}
\end{array} \right),
\qquad
\Gamma_\mu^{(\pm)} = \left(
\begin{array}{c}
\gamma_\mu \gamma_5 \\ \gamma_\mu
\end{array} \right).
\end{align}

The invariant amplitudes arising from the $s$-channel $\Sigma^*$-resonance
exchanges with $J^P=1/2^\pm$, $3/2^\pm$, and $5/2^\pm$ are given by
\begin{align}
\mathcal{M}_{1/2^\pm}^\mu
&= \pm\, i\frac{g_{K N \Sigma^*}\,
g_{\phi \Sigma \Sigma^*}}{s - M_{\Sigma^*}^2 + i M_{\Sigma^*} \Gamma_{\Sigma^*}}
\Gamma^{(\mp)\mu} (\slashed{q}_s + M_{\Sigma^*}) \Gamma^{(\pm)},
\cr
\mathcal{M}_{3/2^\pm}^\mu
&= \mp\, i\frac{g_{K N \Sigma^*}}{M_K} \frac{g_{\phi \Sigma \Sigma^*}}{2M_N}
\frac{1}{s - M_{\Sigma^*}^2 + i M_{\Sigma^*}\Gamma_{\Sigma^*}}
\Gamma_\nu^{(\pm)}
\cr     &
\times (k_2^\alpha g^{\mu\nu} - k_2^\nu g^{\mu\alpha})
(\slashed{q}_s + M_{\Sigma^*}) \Delta_\alpha^\beta(q_s, M_{\Sigma^*})
\Gamma^{(\mp)} k_{1\beta},
\cr
\mathcal{M}_{5/2^\pm}^\mu
&= \pm\, i\frac{g_{K N \Sigma^*}}{M_K^2} \frac{g_{\phi \Sigma \Sigma^*}}{(2M_N)^2}
\frac{1}{s - M_{\Sigma^*}^2 + i M_{\Sigma^*} \Gamma_{\Sigma^*}}
\Gamma_\nu^{(\mp)} k_2^{\alpha_2}
\cr     &
\times (k_2^{\alpha_1} g^{\mu\nu} - k_2^\nu g^{\mu \alpha_1})
(\slashed{q}_s + M_{\Sigma^*}) \Delta_{\alpha_1 \alpha_2}^{\beta_1 \beta_2}
(q_s, M_{\Sigma^*})
\cr   &
\times \Gamma^{(\pm)} k_{1{\beta_1}} k_{1{\beta_2}},
\label{eq:InvAmp2}
\end{align}
with $\mathcal M = \varepsilon_\mu^* \bar{u}_\Sigma \, \mathcal M^\mu \,u_N$ and
$q_s = k_1 + p_1$.
We refer to Ref.~\cite{Kim:2024mqx} for the explicit expressions of the spin-3/2
and spin-5/2 projection operators, $\Delta_{\alpha_1(\alpha_2)}^{\beta_1(\beta_2)}$.
In the present calculation, we set $\Gamma_{\Sigma(2620)}$ = $\Gamma_{\Sigma(3000)}$
= 300 MeV, while $M_{\Sigma(3000)}$ is taken to be 2960 MeV to obtain a better
description of the total cross section.

The following form factors is used for each vertex:
\begin{align}
F_{\Sigma^*} (s) =
\frac{\Lambda_{\Sigma^*}^4}{\Lambda_{\Sigma^*}^4 + (s - M_{\Sigma^*}^2)^2},
\label{eq:FormFac_Res}
\end{align}
where $\Lambda_{\Sigma^*}$ = 1.0 GeV.

\section{Numerical Results}
\label{Sec:III}

\subsection{Born term}
\label{Sec:III-1}

\begin{figure}[ht]
\centering
\includegraphics[width=7.0cm]{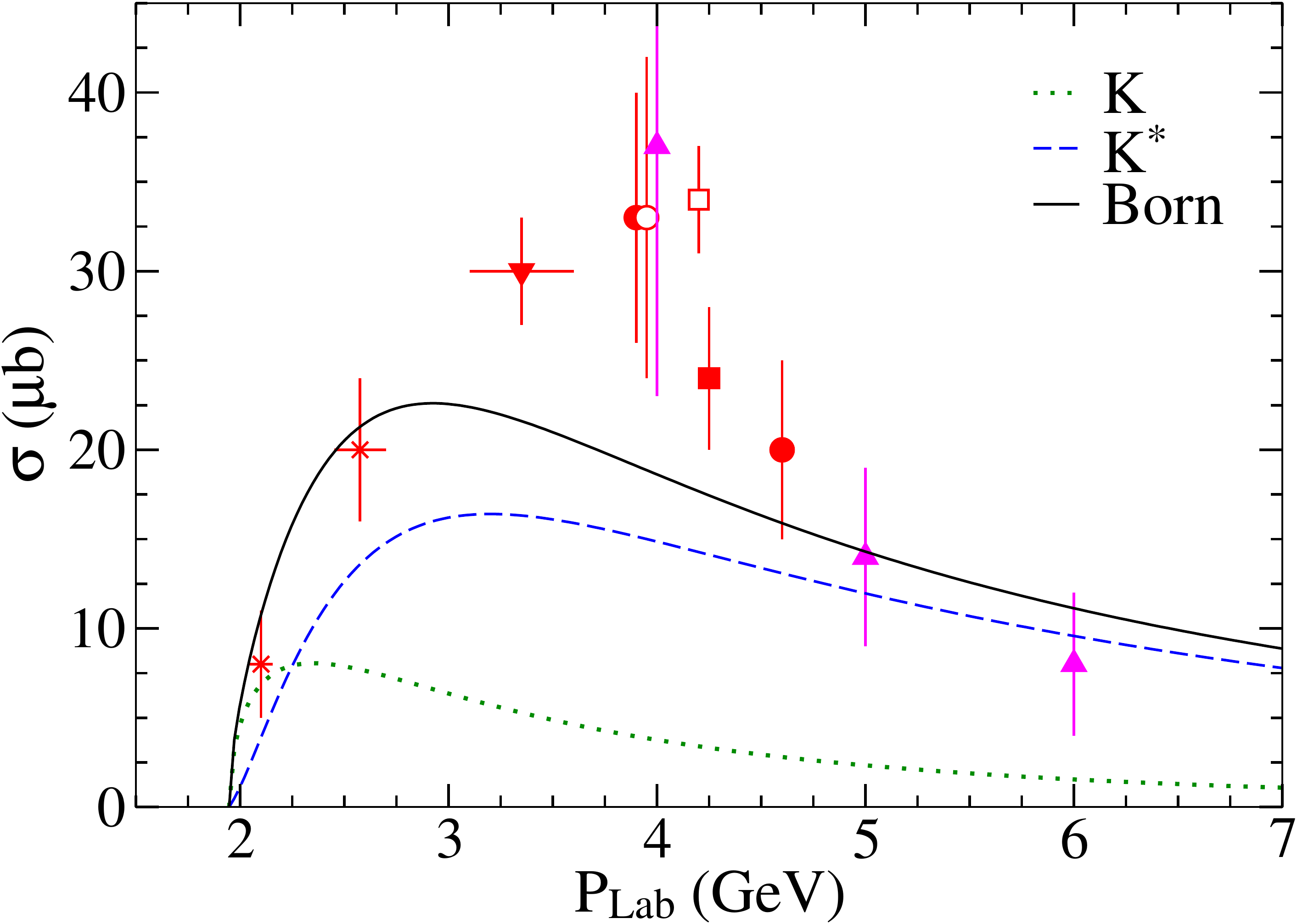}
\caption{Total cross section for $K^- p \to \phi \Sigma^0$ as a function of
$P_{\mathrm{Lab}}$, excluding the $s$-channel $\Sigma^*$-resonance contribution.
The green dotted and blue dashed curves denote the $K$- and $K^*$-Reggeon exchange
contributions, respectively, while the black solid curve represents the Born
contribution.
Experimental data are taken from Refs.~\cite{Lindsey:1966zz} (stars), \cite{
Lyons:1977rp} (filled inverted triangles), \cite{Aguilar-Benitez:1972ngz} (filled
circles), \cite{Rouge:1972xr} (open circles), \cite{Ayres:1974aj} (filled triangles),
\cite{ACNO:1977ctz} (open squares), and \cite{DeGroot:1974qw} (filled squares).}
\label{FIG02}
\end{figure}
We first present our numerical results for the $K^- p \to \phi \Sigma^0$ reaction
by considering only the Born contributions, i.e., the nonresonant background terms.
These include the Reggeized $t$-channel exchanges as well as the $s$- and
$u$-channel ground-state baryon exchanges discussed in Sec.~\ref{Sec:II-1}.
Figure~\ref{FIG02} shows the total cross section as a function of $P_{\mathrm{Lab}}$,
together with the individual contributions from the $K$- and $K^*$-Reggeon exchanges.
The $K$-Reggeon exchange exhibits a sharp rise, whereas the $K^*$-Reggeon exchange
shows a more gradual increase, reaches a peak at $P_{\mathrm{Lab}} \simeq 2.8\,
\mathrm{GeV}$, and then decreases at higher beam momenta.
The $K^*$ exchange dominates over the $K$ exchange in magnitude.
In the region $3.0 \leqslant P_{\mathrm{Lab}} \leqslant 4.5\, \mathrm{GeV}$, the
coherent sum of the two exchanges shows a significant deviation from the experimental
data~\cite{Lindsey:1966zz,Lyons:1977rp,Aguilar-Benitez:1972ngz,Rouge:1972xr,
Ayres:1974aj,ACNO:1977ctz,DeGroot:1974qw}, suggesting the need for additional
reaction mechanisms.
On the other hand, the $s$-channel $\Sigma$-exchange and $u$-channel
nucleon-exchange contributions are negligible and are therefore not shown.

\begin{figure}[ht]
\centering
\includegraphics[width=8.5cm]{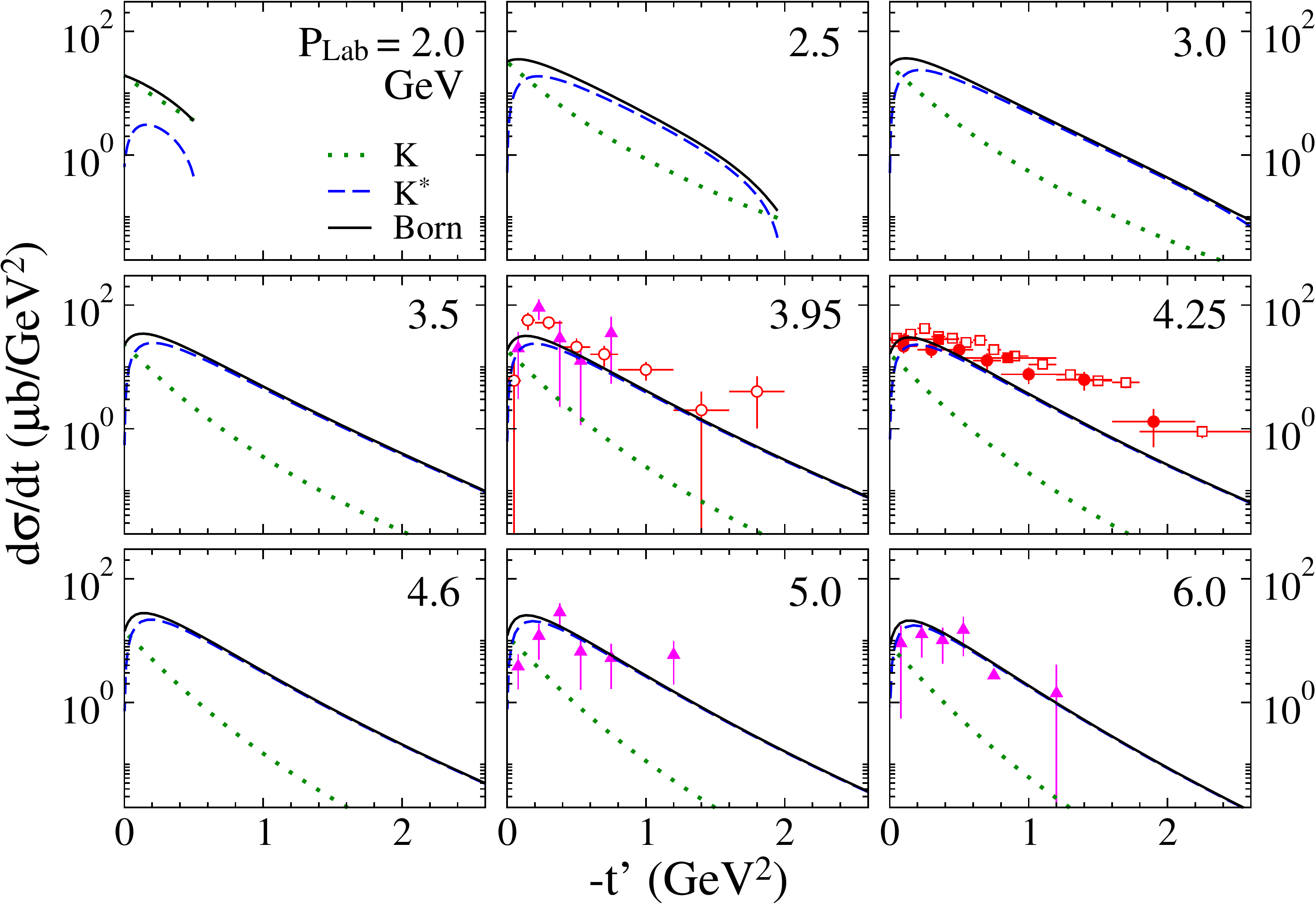}
\caption{$t$-dependent differential cross sections for $K^- p \to \phi \Sigma^0$,
excluding the $s$-channel $\Sigma^*$-resonance contribution, at nine fixed beam
momenta.
Experimental data are taken from Refs.~\cite{Aguilar-Benitez:1972ngz} (filled
circles), \cite{Rouge:1972xr} (open circles), \cite{Ayres:1974aj} (filled triangles),
\cite{ACNO:1977ctz} (open squares), and \cite{DeGroot:1974qw} (filled squares).
The curve notations are the same as in Fig.~\ref{FIG02}.}
\label{FIG03}
\end{figure}

We display the differential cross sections as functions of $-t' = -t + t_{min}$ in
Fig.~\ref{FIG03} at nine fixed beam momenta.
In the forward-angle limit, the two Reggeon exchanges lead to quite different angular
dependences.
The available experimental data strongly favor the $K^*$-Reggeon exchange over the
$K$-Reggeon exchange~\cite{Aguilar-Benitez:1972ngz,Rouge:1972xr,Ayres:1974aj,
ACNO:1977ctz,DeGroot:1974qw}.
However, as $-t'$ increases, the Born results begin to deviate increasingly from the
experimental data, as is clearly seen at $P_\mathrm{Lab} = 4.25\,\mathrm{GeV}$.
This behavior suggests that an additional mechanism, other than the $t$-channel
process, is needed to account for the experimental data in the region 
$3.0 \leqslant P_{\mathrm{Lab}} \leqslant 4.5\, \mathrm{GeV}$.
This mechanism will be discussed in detail in Sec.~\ref{Sec:III-2}.

We also examine the SDMEs, which provide key information on the spin structure and
underlying reaction mechanism~\cite{Kim:2017hhm}.
In this work, we focus on the polarization of the produced vector meson only.
The case where the recoil $\Sigma^0$ baryon is also polarized is discussed
in Ref.~\cite{Kim:2017hhm}.
We consider the decay channel $\phi \to K^+ K^-$ in the $K^- p \to \phi \Sigma^0$
reaction.
The analysis of the outgoing $K^+$ angular distribution in the vector-meson rest
frame involves an ambiguity in the choice of the quantization axis.
One possible choice is to take this axis to be antiparallel to the momentum of the
outgoing hyperon $\Sigma^0$.
Alternatively, it can be chosen parallel to the momentum of the incoming meson.
We refer to the former choice as the helicity (H) frame and to the latter as the
Gottfried--Jackson (GJ) frame, following the convention of Refs.~\cite{
Crennell:1972km,Schilling:1969um}.
The helicity frame is commonly used to test $s$-channel helicity conservation, while
the GJ frame is more suitable for examining the $t$-channel exchange mechanism.

The decay distribution is described in terms of the SDMEs $\rho_{\lambda \lambda'}$,
where the vector-meson helicity $\lambda_V$ is denoted by $\lambda$.
The SDMEs are constructed from the full amplitude $T$, given by the coherent
sum of Eqs.~(\ref{eq:RegAmpl1}), (\ref{eq:Ampl2}), and (\ref{eq:InvAmp2}), as
\begin{align}
\rho_{\lambda\lambda'} = \frac{1}{{\mathcal N}^2}
\sum\limits_{s_f = \pm\frac12,\, s_i = \pm\frac12}
T_{\lambda,s_f;s_i}\, T^*_{\lambda',s_f;s_i},
\label{eq:SDME}
\end{align}
where the normalization factor is given by
\begin{align}
{\mathcal N}^2 = \sum_{\lambda,s_f,s_i} |T_{\lambda,s_f ; s_i}|^2.
\label{eq:SDME-NF}
\end{align}
The SDMEs satisfy the Hermiticity and parity-symmetry relations
\begin{align}
\rho_{\lambda \lambda'} = \rho_{\lambda' \lambda}^*,\,\,\,
\rho_{-\lambda, -\lambda'} = (-1)^{\lambda - \lambda'} \rho_{\lambda \lambda'}.
\label{eq:SDME1}
\end{align}
Together with the normalization condition
\begin{align}
\rho_{00} + \rho_{11} + \rho_{-1-1} = 1,
\label{eq:SDME2}
\end{align}
these relations are satisfied in our numerical calculations.

\begin{figure}[ht]
\centering
\includegraphics[width=8.5cm]{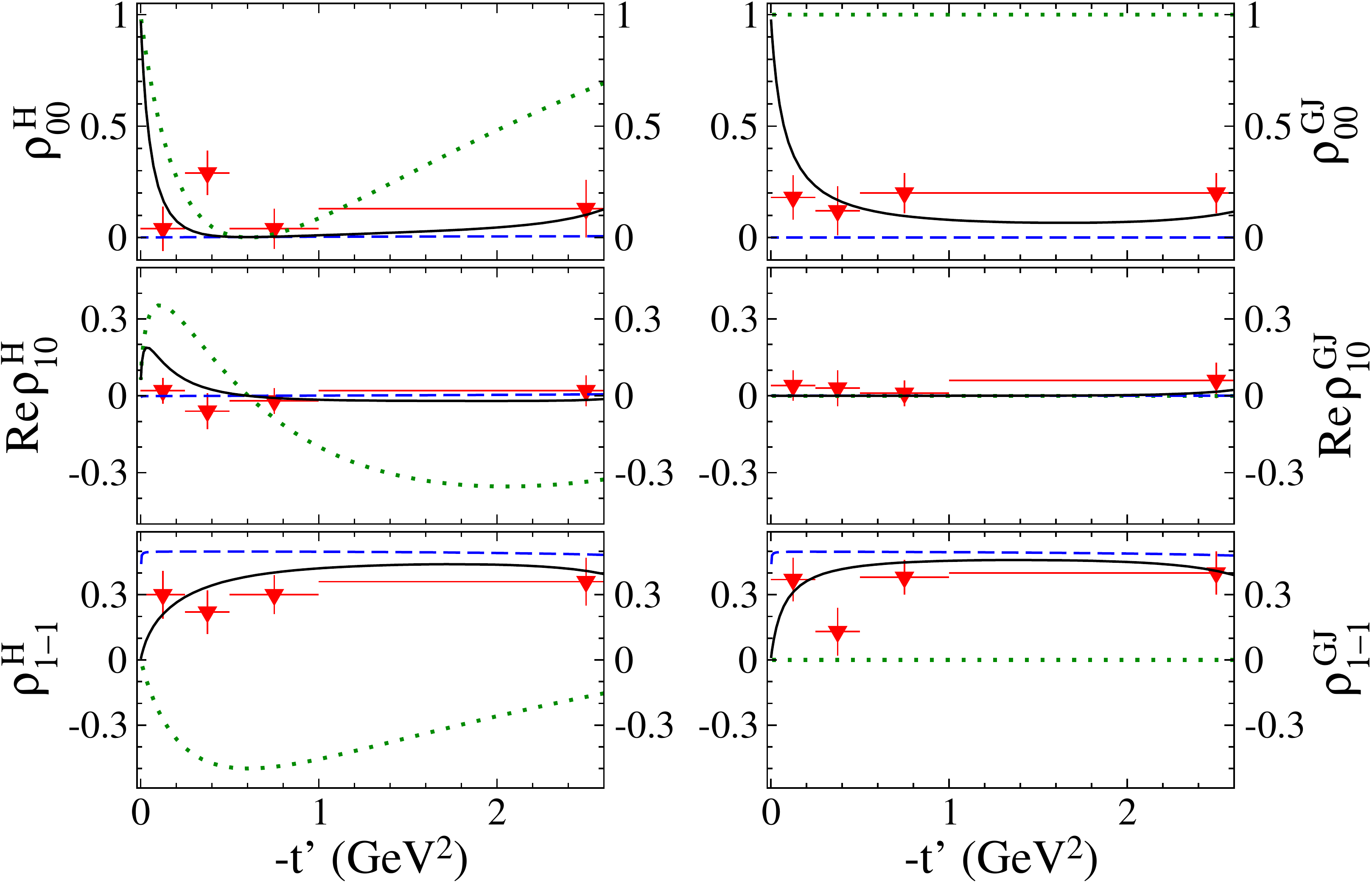}
\caption{SDMEs $\rho_{00}$, Re\,$\rho_{10}$, and $\rho_{1-1}$ for $K^- p
\to \phi \Sigma^0$ as functions of $-t'$ in the helicity and GJ frames at
$P_{\mathrm{Lab}}$ = 3.35 GeV.
Experimental data are taken from Ref.~\cite{Lyons:1977rp}.
The curve notations are the same as in Fig.~\ref{FIG02}.}
\label{FIG04}
\end{figure}
\begin{figure*}[ht]
\centering
\includegraphics[width=8.5cm]{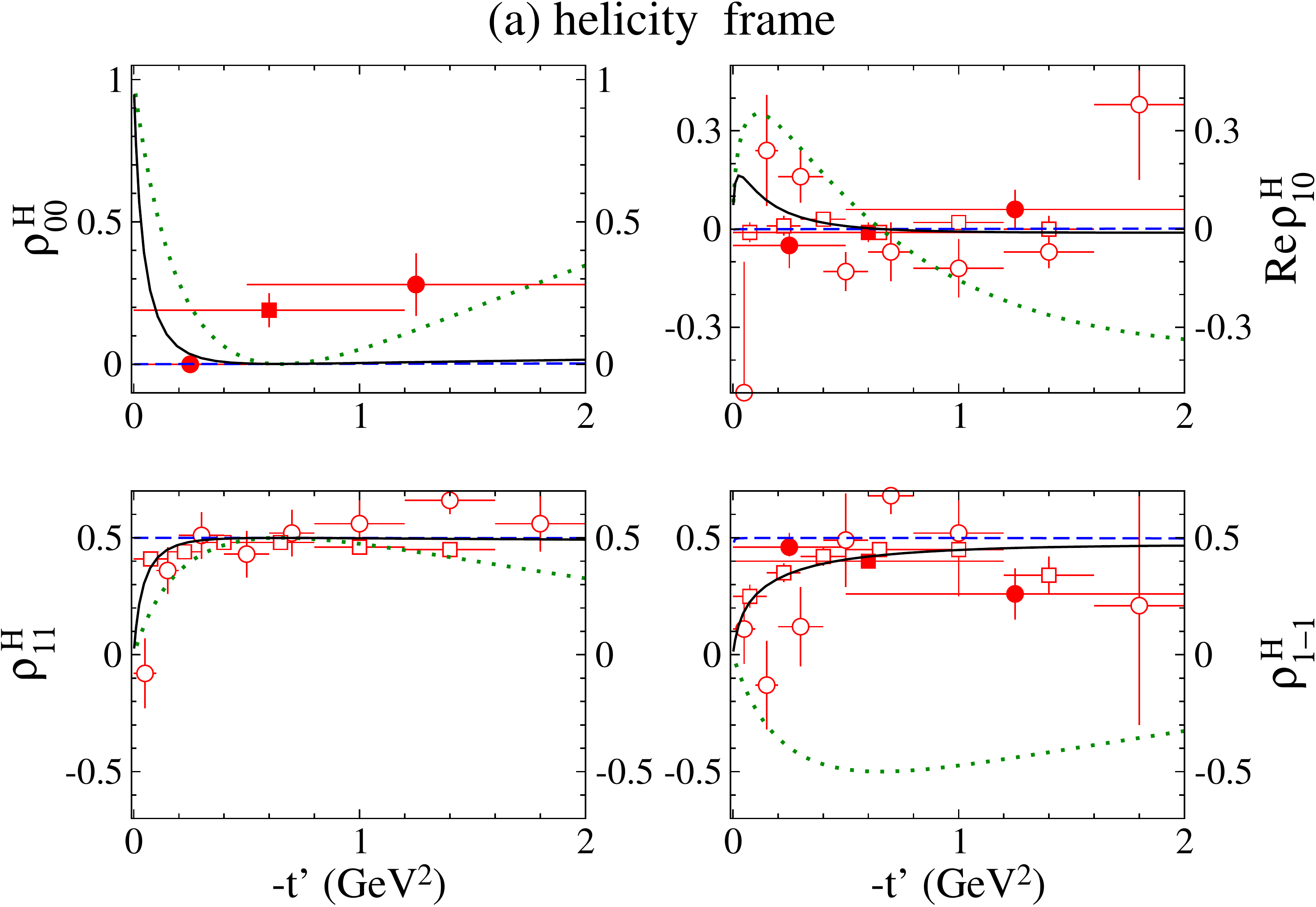} \hspace{1.0em}
\includegraphics[width=8.5cm]{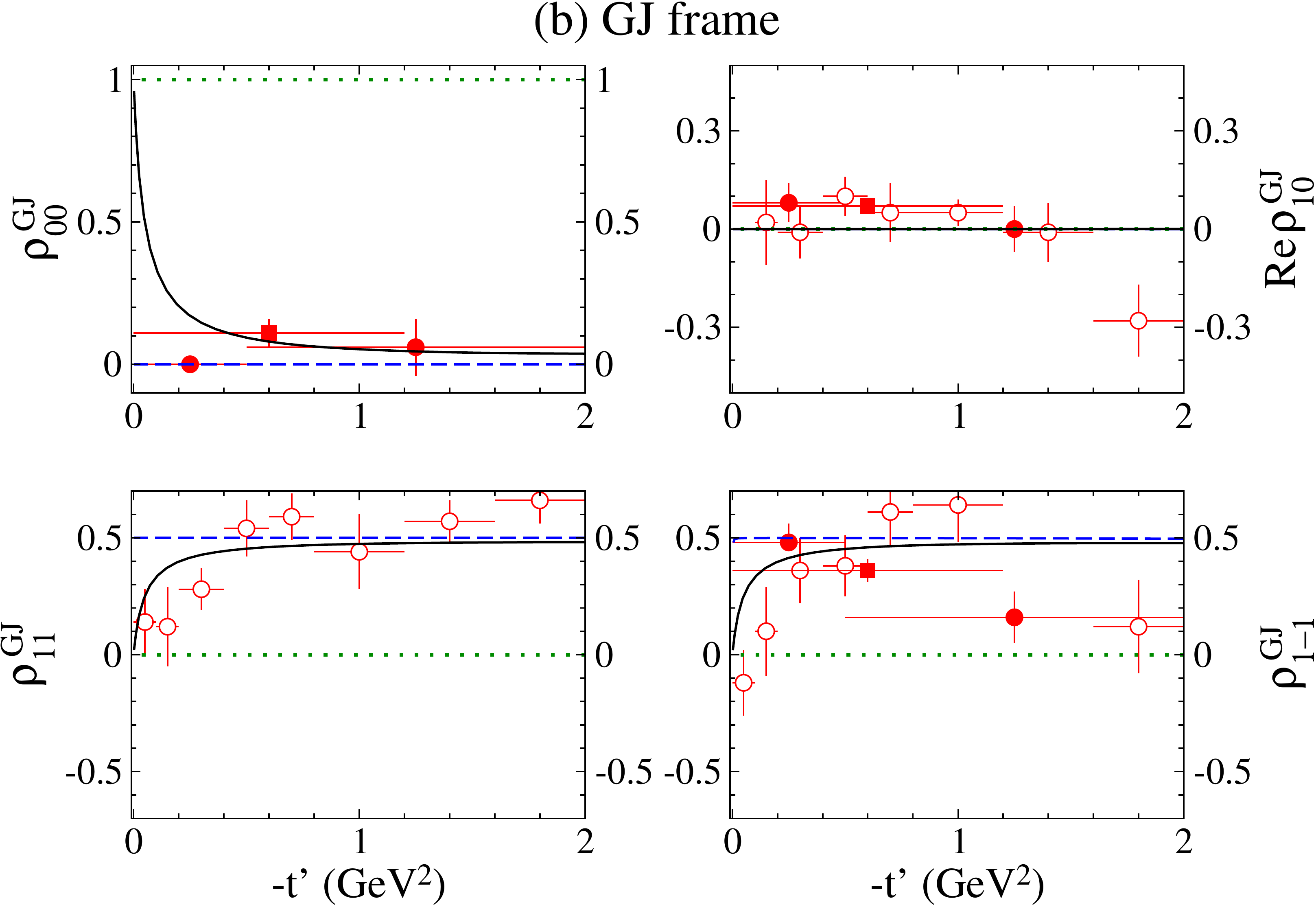}
\caption{SDMEs $\rho_{00}$, Re\,$\rho_{10}$, $\rho_{11}$, and $\rho_{1-1}$ for
$K^- p \to \phi \Sigma^0$ as functions of $-t'$ in the (a) helicity and (b) GJ
frames at $P_{\mathrm{Lab}}$ = 4.25 GeV.
Experimental data are taken from Refs.~\cite{Aguilar-Benitez:1972ngz} (filled
circles), \cite{Rouge:1972xr} (open circles), \cite{ACNO:1977ctz} (open squares),
and \cite{DeGroot:1974qw} (filled squares).
The beam momentum in Ref.~\cite{Rouge:1972xr} is $P_{\mathrm{Lab}}$ = 3.95 GeV.
The curve notations are the same as in Fig.~\ref{FIG02}.}
\label{FIG05}
\end{figure*}

With these SDMEs, the decay angular distribution is written as
\begin{align}
W^{0}(\Omega_f) &= \frac{3}{4\pi} \Bigl[ \rho_{00}\cos^2\Theta +
\rho_{11}\sin^2\Theta - \rho_{1-1} \sin^2\Theta \cos2\Phi
\cr &
- {\sqrt{2}}\, \mathrm{Re}(\rho_{10}) \sin2\Theta\cos\Phi \Bigr],
\label{eq:DAD}
\end{align}
where $\Theta$ and $\Phi$ denote the polar and the azimuthal angles of the outgoing
$K^+$ meson, respectively~\cite{Kim:2017hhm}.

In Fig.~\ref{FIG04}, we present the results for the SDMEs $\rho_{00}$,
Re\,$\rho_{10}$, and $\rho_{1-1}$ as functions of $-t'$ in the helicity and GJ frames
at $P_{\mathrm{Lab}} = 3.35\,\mathrm{GeV}$~\cite{Lyons:1977rp}.
Figure~\ref{FIG05} also shows the results for $\rho_{00}$, Re\,$\rho_{10}$,
$\rho_{11}$, and $\rho_{1-1}$ at $P_{\mathrm{Lab}} = 4.25\,\mathrm{GeV}$~\cite{
Aguilar-Benitez:1972ngz,Rouge:1972xr,ACNO:1977ctz,DeGroot:1974qw}.
The individual $K$- and $K^*$-Reggeon contributions exhibit distinctly different
behaviors.
For the vector-meson Reggeon exchange, $\rho_{\lambda\lambda'}$ with
$|\lambda| = |\lambda'| = 1$ are selectively enhanced because of the spin
structure $\epsilon^{\mu\nu\alpha\beta} \varepsilon^*_\mu (\lambda_V) k_{2\alpha}
k_{1\beta}$ of the amplitude in Eq.~(\ref{eq:Ampl1}).
This factor is proportional to the vector product $\bm{\varepsilon}^* (\lambda_V)
\times \bm{k}_1$ in the vector meson rest frame.
In the helicity frame, at small $-t'$, the incident $K^-$ three-momentum $\bm{k}_1$
is almost parallel to the $z$ axis, so that $\bm{\varepsilon}^*(\lambda_V) \times
\bm{k}_1 \simeq i\lambda_V \bm{\varepsilon}^*(\lambda_V)|\bm{k}_1|$.
This kinematic feature enhances the SDMEs with $|\lambda|=|\lambda^\prime|=1$.
In the GJ frame, $\bm{k}_1$ defines the quantization axis, and hence the SDMEs
vanish when either $\lambda = 0$ or $\lambda' = 0$.
By contrast, for the pseudoscalar-Reggeon exchange, the amplitude is proportional
to the scalar product $\bm{\varepsilon}^*(\lambda_V) \cdot \bm{k}_1$ in
Eq.~(\ref{eq:Ampl1}), which selects the helicity-zero component in the GJ frame.
Consequently, $\rho_{00}=1$, while all other $\rho_{\lambda\lambda'}$
vanish~\cite{Kim:2017hhm}.

Overall, the coherent sum of the two exchange contributions is in good agreement
with the experimental data.
Some discrepancies remain for the SDMEs other than $\rho_{11}$ in the region
$-t' \geqslant 1\, \mathrm{GeV}^2$ at $P_{\mathrm{Lab}} = 4.25\, \mathrm{GeV}$ as
shown in Fig.~\ref{FIG05}; these will be discussed in detail in the next subsection.
Note that the diagonal element $\rho_{00}$ reflects the helicity-zero
unnatural-parity contribution, which arises mainly from the $K$-Reggeon exchange.
In the transverse sector with $\lambda_V = \pm 1$, the combinations $\rho_{11} \pm
\rho_{1-1}$ distinguish the natural- and unnatural-parity contributions, associated
primarily with the $K^*$- and $K$-Reggeon exchanges, respectively.
The results in Figs.~\ref{FIG04} and \ref{FIG05}, together with the differential
cross sections in Fig.~\ref{FIG03}, show that the reaction mechanism is dominated
by natural-parity exchange.
They also indicate that additional exchanges, such as the axial-vector $K_1(1270)$
and $K_1(1400)$ mesons or the scalar $\kappa$ meson [$K_0^*(700)$], are not required
to describe the present data.

\subsection{Resonance term}
\label{Sec:III-2}

With the Born-term contribution fixed, we additionally include the $\Sigma(2620)$ and
$\Sigma(3000)$ resonances in the $s$ channel to describe the local structures
observed in the region $3.0 \leqslant P_{\mathrm{Lab}} \leqslant 4.5\, \mathrm{GeV}$
as discussed in Sec.~\ref{Sec:II-2}.
Because the spin and parity of these resonances have not yet been firmly established,
we consider the assignments $J^P = 1/2^\pm$, $3/2^\pm$, and $5/2^\pm$.

No experimental information is currently available on the $\Sigma^* \to \bar{K} N$
and $\Sigma^* \to \phi \Sigma$ branching ratios for either the $\Sigma(2620)$ or
$\Sigma(3000)$ resonance.
We therefore treat the relevant coupling products as free phenomenological parameters
and adjust them to describe the overall behavior of the available data.
The products of the coupling constants, $g_{K N \Sigma^*} \cdot
g_{\phi \Sigma \Sigma^*}$, appearing in Eqs.~(\ref{eq:ResLag1}) and
(\ref{eq:ResLag2}), and the corresponding products of branching ratios are summarized
in Table~\ref{TAB:1}.
Here, $\mathcal{B}_{\bar{K} N} \equiv \mathcal{B}_{\Sigma^*\to \bar{K} N}$ and
$\mathcal{B}_{\phi \Sigma} \equiv \mathcal{B}_{\Sigma^* \to \phi \Sigma}$, where
$\Sigma^*$ denotes the resonance listed in the first column.
\begin{table}[ht]
\caption{The products of the coupling constants and the corresponding products of
branching ratios for the $\Sigma(2620)$ and $\Sigma(3000)$ resonances.}
\begin{tabular}{cccc}
\hline\hline
State & \hspace{1em}$J^P$\hspace{1em}
& \hspace{1em}$g_{K N \Sigma^*} \cdot g_{\phi \Sigma \Sigma^*}$\hspace{1em}
&$\mathcal{B}_{\bar{K} N} \cdot \mathcal{B}_{\phi \Sigma}$ \\
\hline
$\Sigma(2620)$
& $1/2^+$ & 2.37 & $2.2 \cdot 10^{-3}$  \\
& $1/2^-$ & 2.63 & $1.3 \cdot 10^{-3}$  \\
& $3/2^+$ & 0.16 & $6.5 \cdot 10^{-4}$  \\
& $3/2^-$ & 0.22 & $6.7 \cdot 10^{-4}$  \\
& $5/2^+$ & 0.60 & $4.2 \cdot 10^{-4}$  \\
& $5/2^-$ & 0.35 & $2.3 \cdot 10^{-4}$  \\
&   \\
$\Sigma(3000)$
& $1/2^+$ & 4.02 & $1.0 \cdot 10^{-2}$  \\
& $1/2^-$ & 5.76 & $8.6 \cdot 10^{-3}$  \\
& $3/2^+$ & 0.18 & $3.9 \cdot 10^{-3}$  \\
& $3/2^-$ & 0.28 & $4.3 \cdot 10^{-3}$  \\
& $5/2^+$ & 0.50 & $3.3 \cdot 10^{-3}$  \\
& $5/2^-$ & 0.33 & $3.0 \cdot 10^{-3}$  \\
\hline\hline
\end{tabular}
\label{TAB:1}
\end{table}

\begin{figure*}[ht]
\centering
\includegraphics[width=7.0cm]{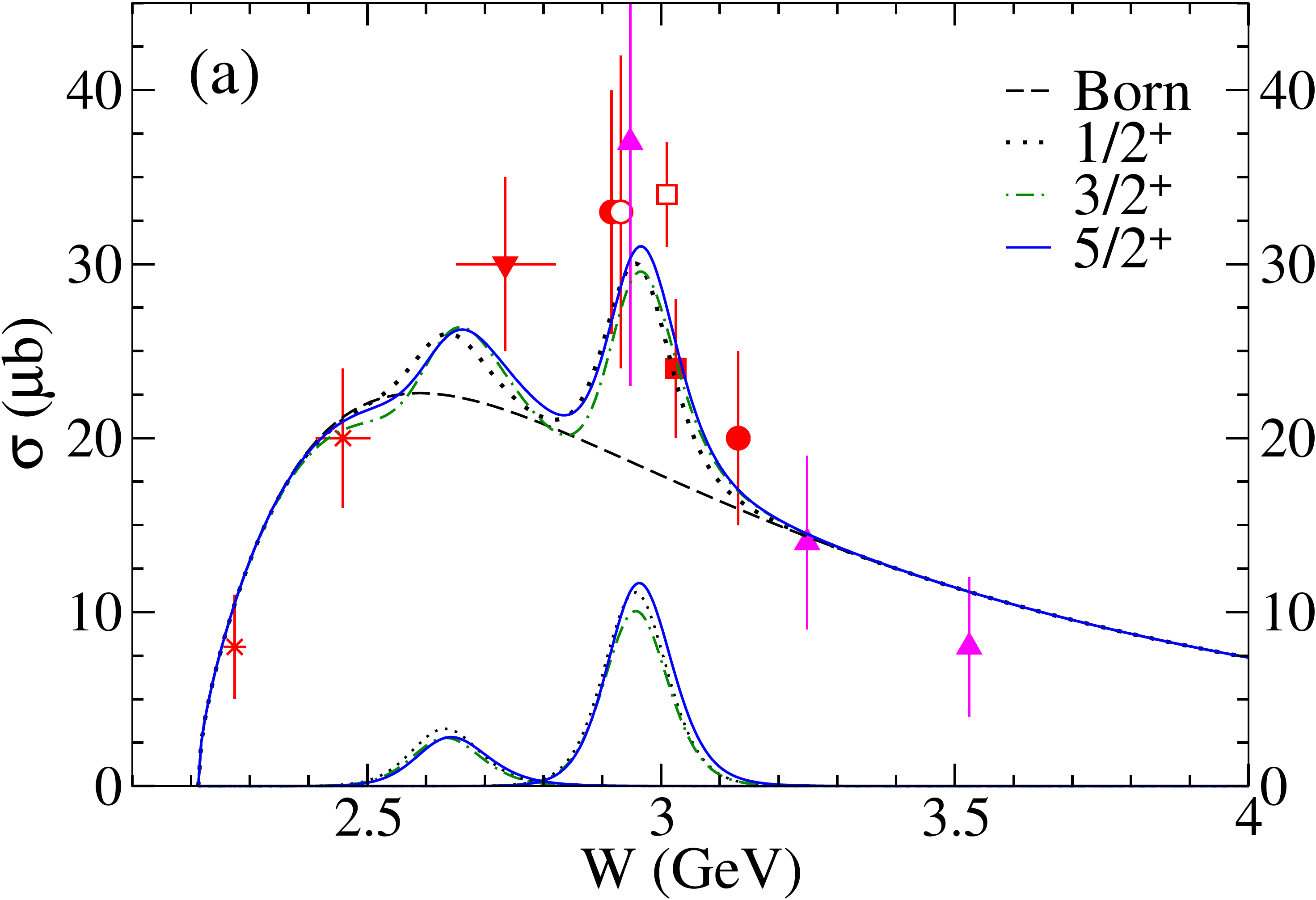} \hspace{1.5em}
\includegraphics[width=7.0cm]{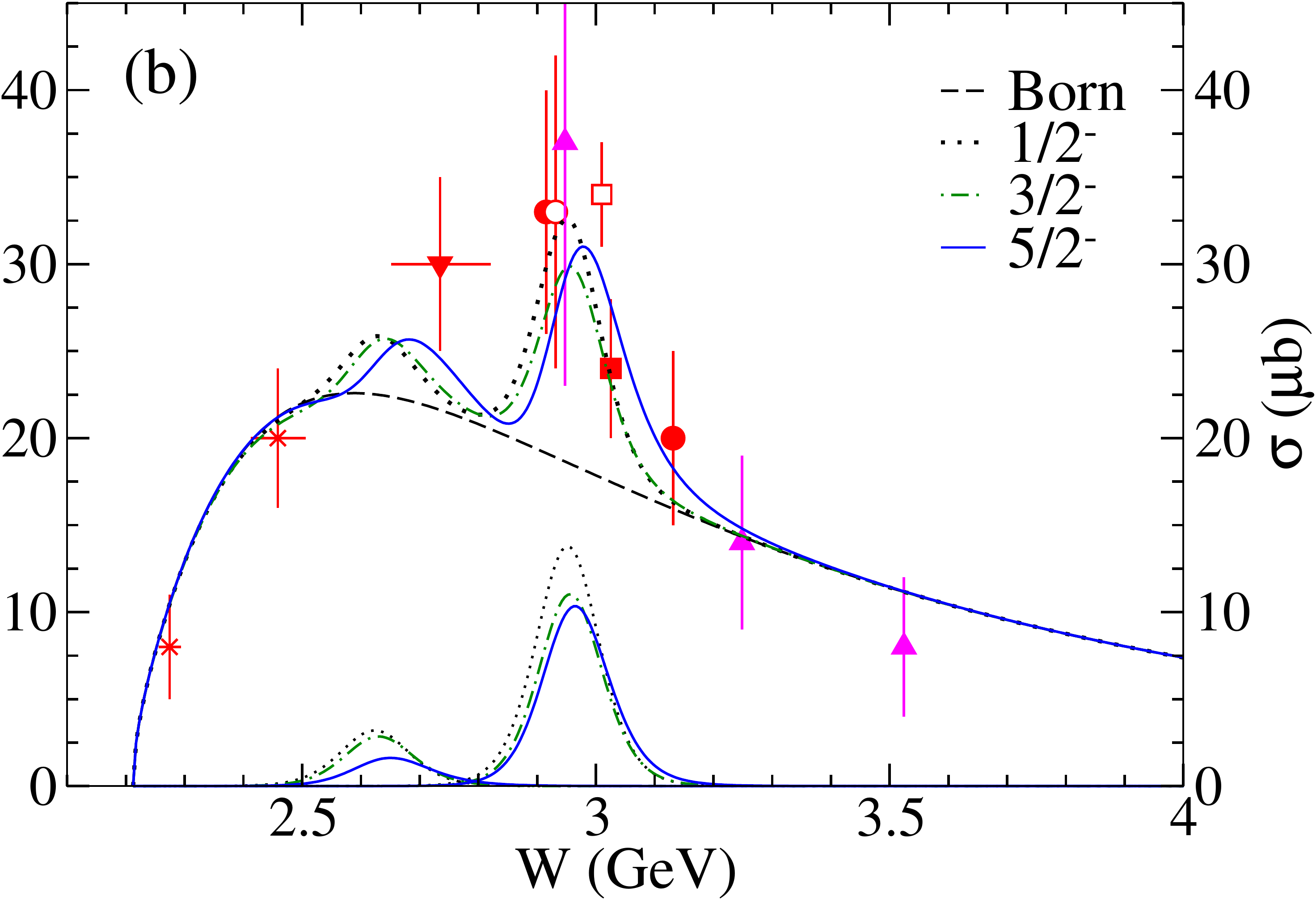}
\caption{Total cross sections for $K^- p \to \phi \Sigma^0$ as functions of $W$.
The black dashed curve denotes the Born-term contribution, while the other curves
show the full results obtained by including the $s$-channel $\Sigma^*$-resonance
contributions.
The individual $\Sigma(2620)$ and $\Sigma(3000)$ contributions are also shown
separately.
Panels (a) and (b) correspond to the $J^P$ = $(1/2^+,\, 3/2^+,\, 5/2^+)$ and $J^P$
= $(1/2^-,\, 3/2^-,\, 5/2^-)$ assignments for the $\Sigma^*$ resonances,
respectively.
Experimental data are shown with the same symbols as in Fig.~\ref{FIG02}.}
\label{FIG06}
\end{figure*}
\begin{figure*}[ht]
\centering
\includegraphics[width=8.5cm]{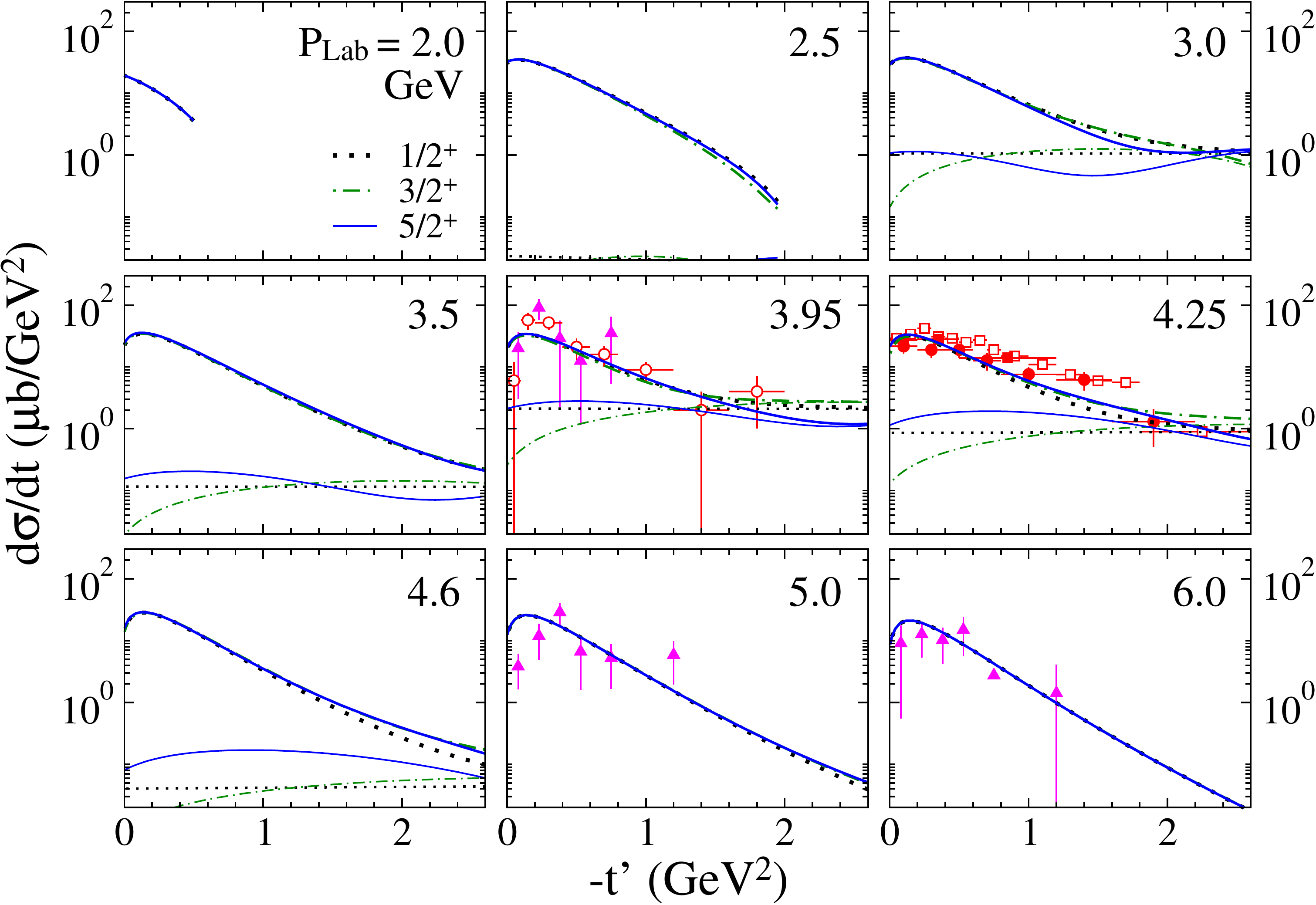} \hspace{0.5em}
\includegraphics[width=8.5cm]{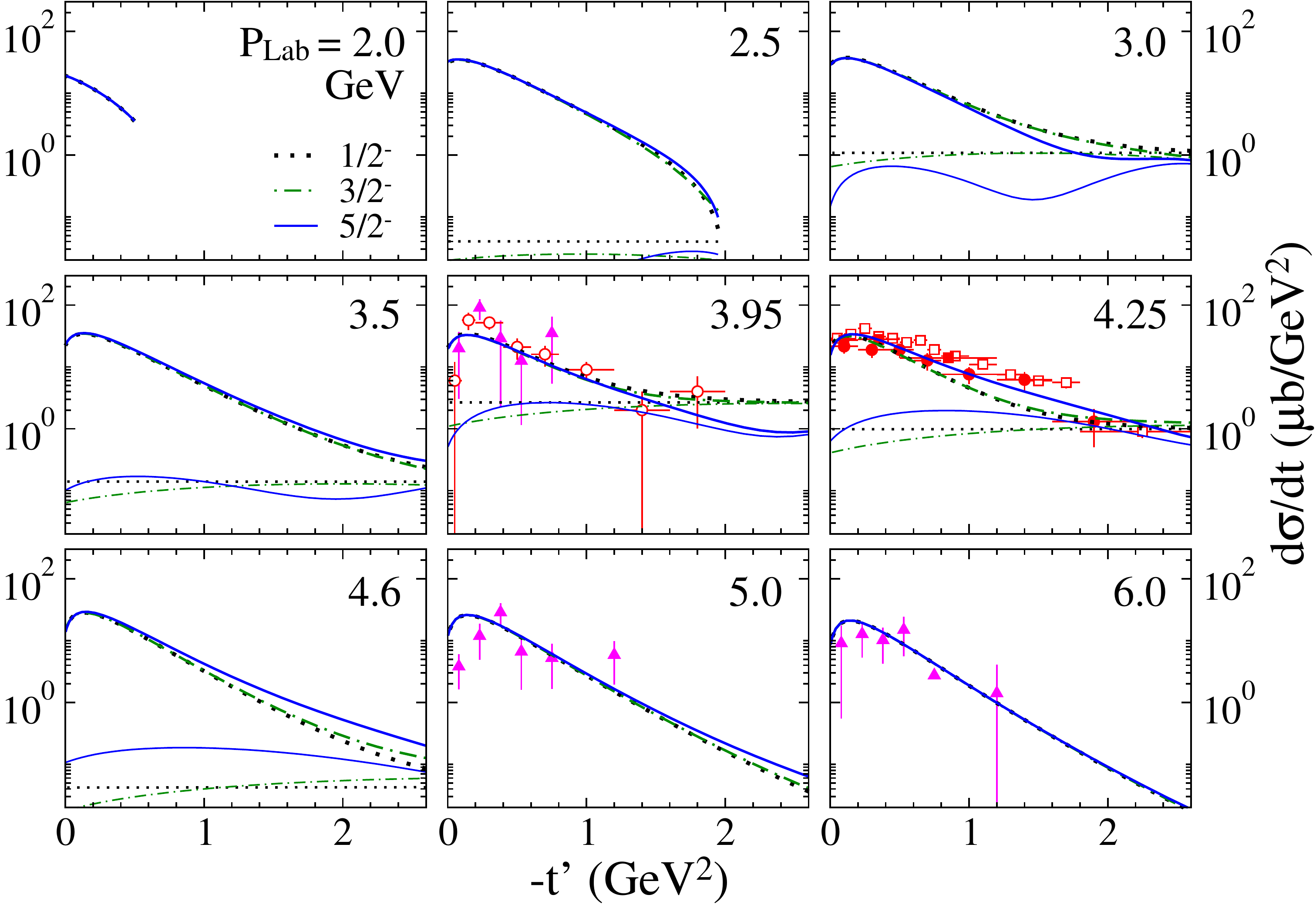}
\caption{$t$-dependent differential cross sections for $K^- p \to \phi \Sigma^0$,
including the $s$-channel $\Sigma^*$-resonance contributions, at nine fixed beam
momenta.
The individual $\Sigma(2620)$ and $\Sigma(3000)$ contributions are also shown
separately at $2.5 \leqslant P_{\rm Lab} \leqslant 4.6\, \mathrm{GeV}$.
Left and right panels correspond to the $J^P$ = $(1/2^+,\, 3/2^+,\, 5/2^+)$ and
$J^P$ = $(1/2^-,\, 3/2^-,\, 5/2^-)$ assignments for the $\Sigma^*$ resonances,
respectively.
Experimental data are shown with the same symbols as in Fig.~\ref{FIG03}.}
\label{FIG07}
\end{figure*}
\begin{figure*}[ht]
\centering
\includegraphics[width=8.5cm]{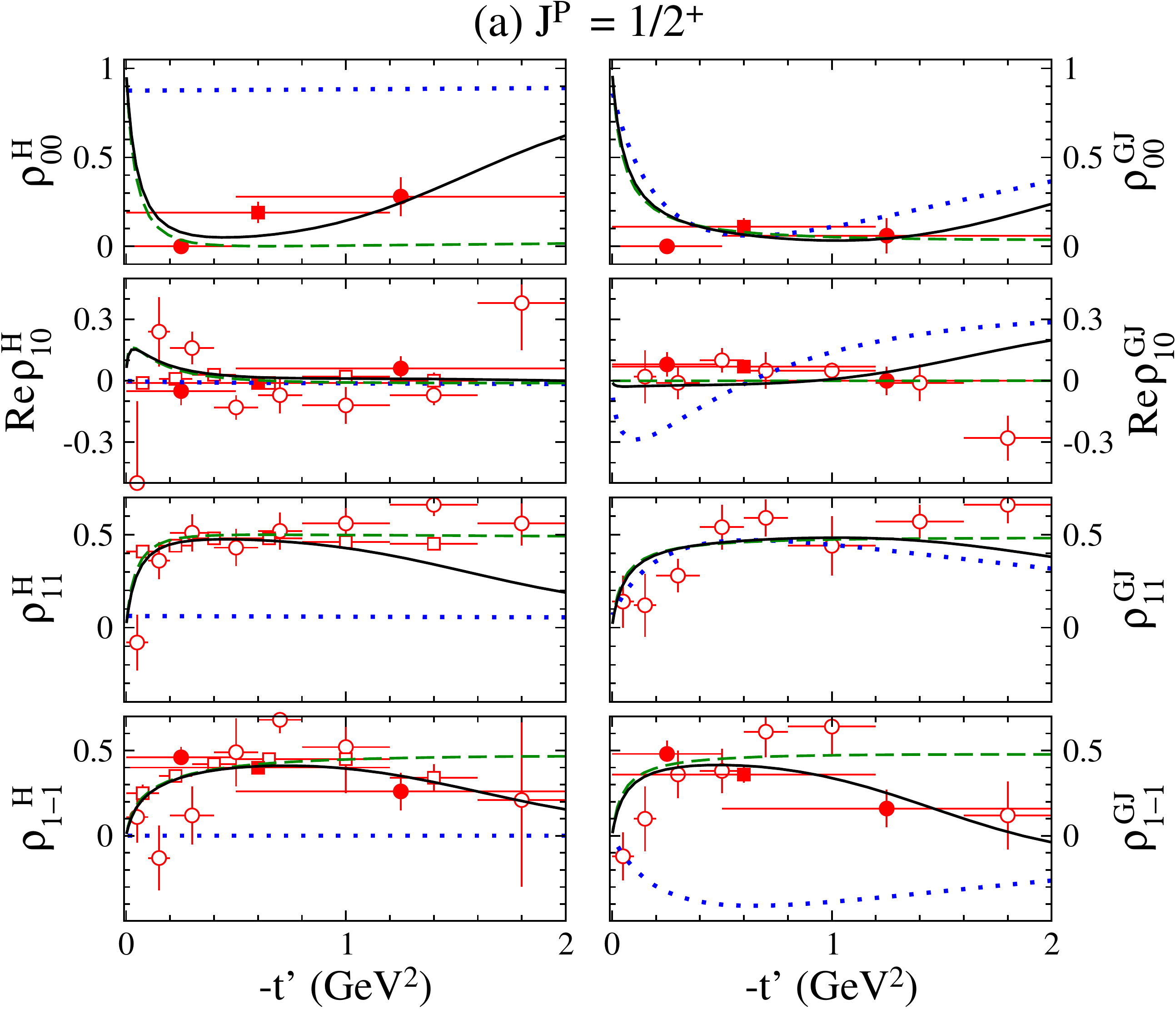} \hspace{1.5em}\vspace{1.0em}
\includegraphics[width=8.5cm]{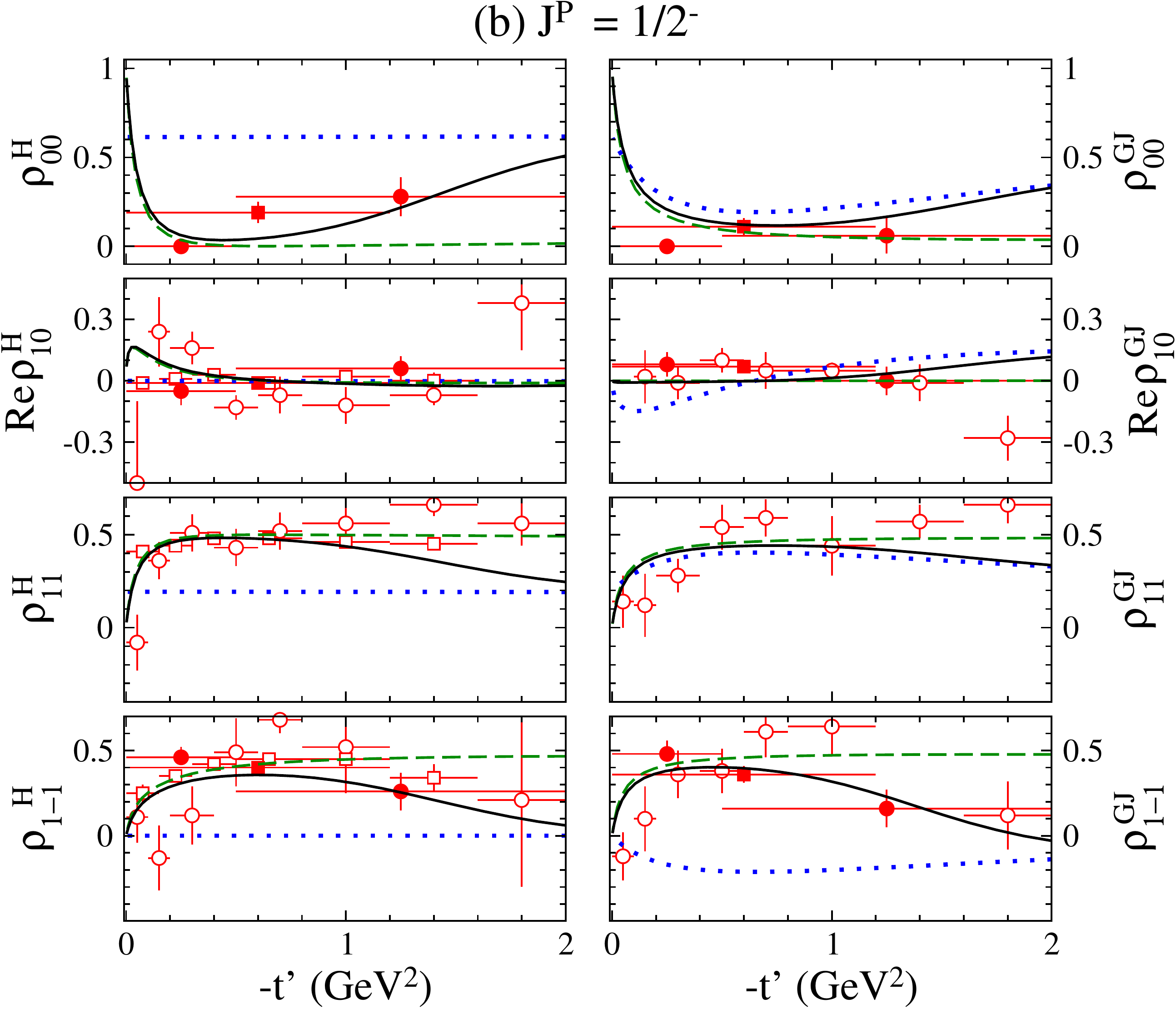}
\includegraphics[width=8.5cm]{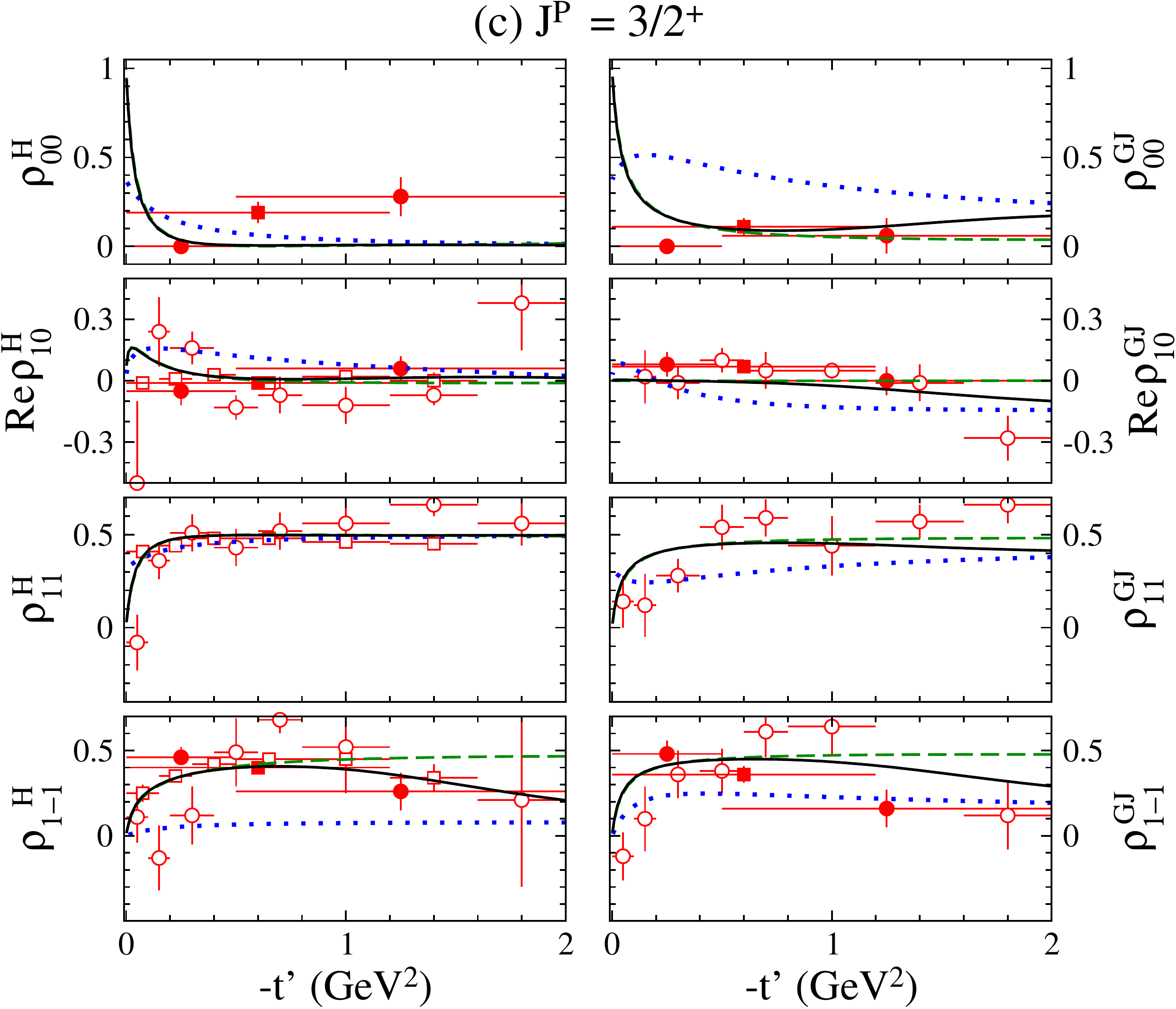} \hspace{1.5em}\vspace{1.0em}
\includegraphics[width=8.5cm]{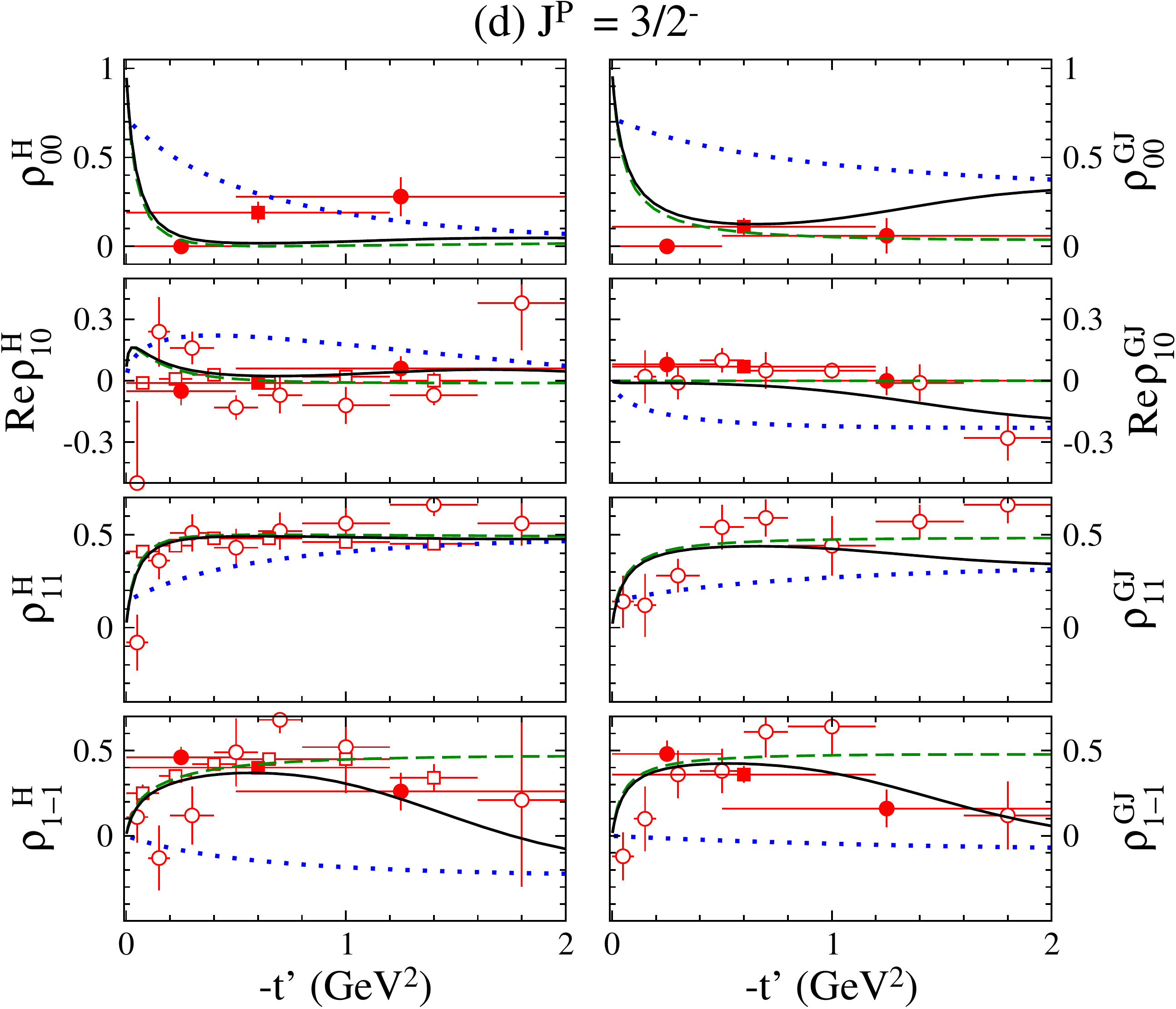}
\includegraphics[width=8.5cm]{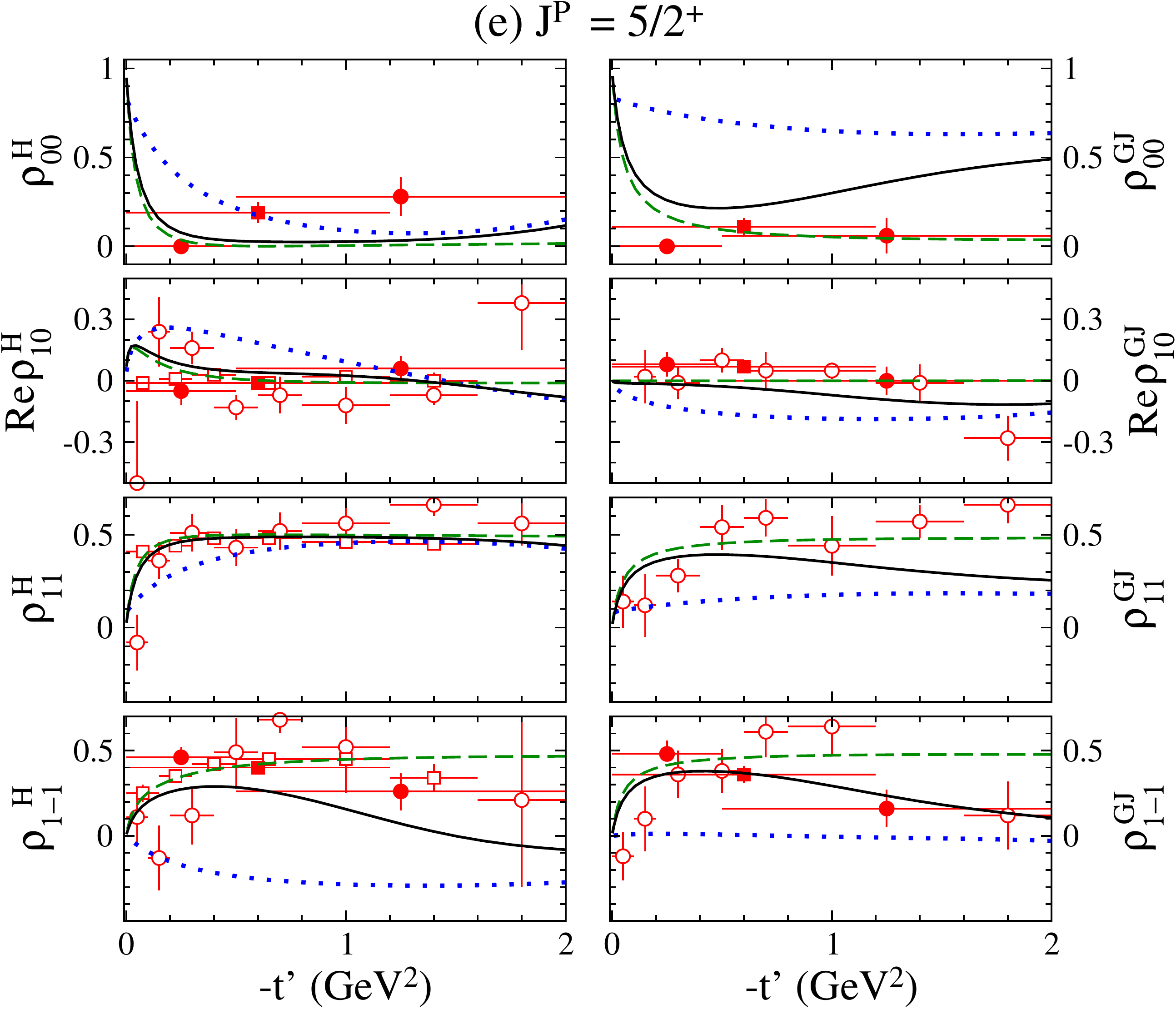} \hspace{1.5em}
\includegraphics[width=8.5cm]{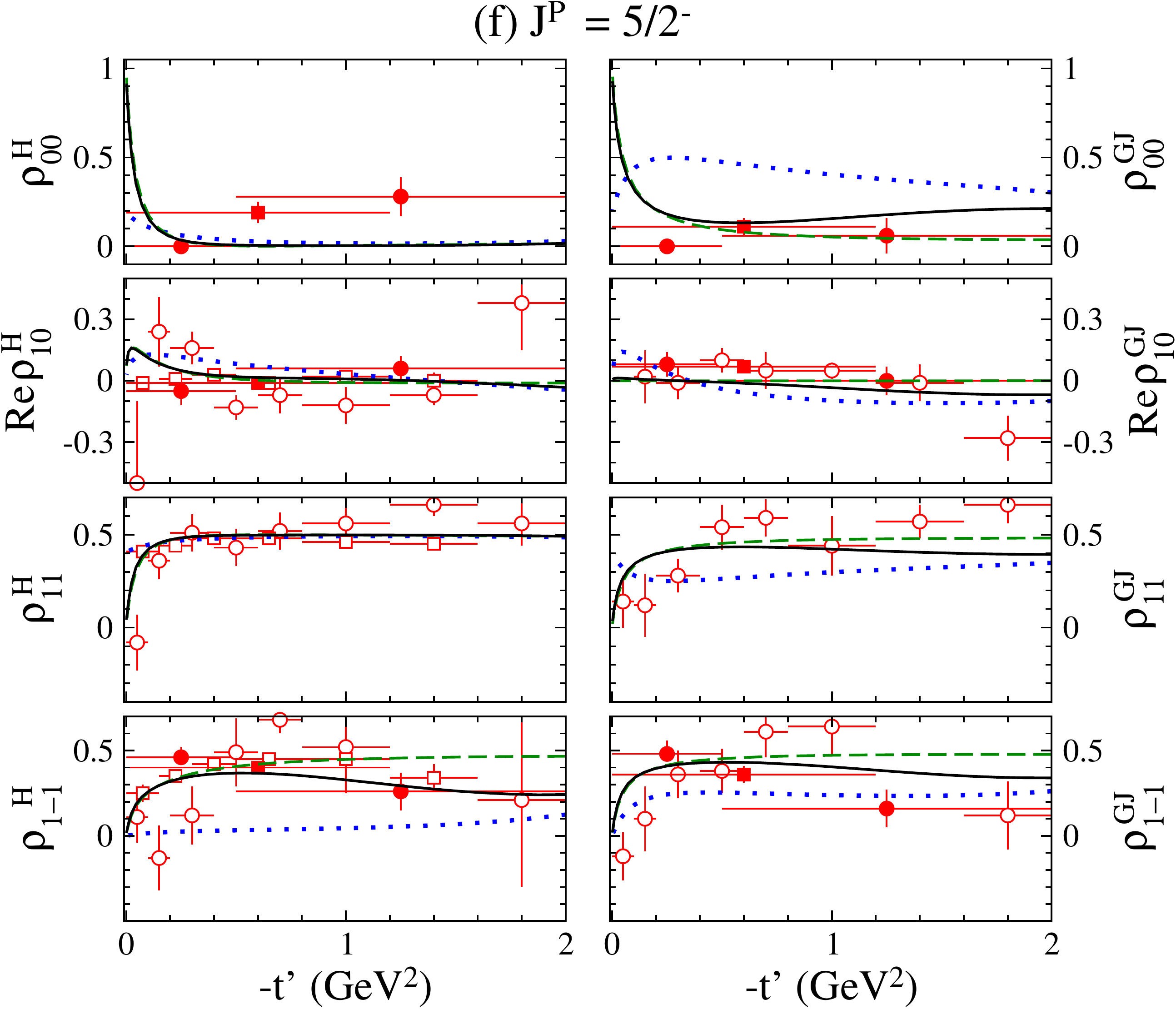}
\caption{SDMEs $\rho_{00}$, Re\,$\rho_{10}$, $\rho_{11}$, and $\rho_{1-1}$
for $K^- p \to \phi \Sigma^0$ in the helicity and GJ frames  at $P_{\mathrm{Lab}} =
4.25\, \mathrm{GeV}$.
The green dashed, blue dotted, and black solid curves denote the Born-term
contribution, $\Sigma(3000)$ contribution, and full result, respectively.
Panels (a)-(f) correspond to the $J^P = 1/2^+$, $1/2^-$, $3/2^+$, $3/2^-$, $5/2^+$,
and $5/2^-$ assignments for the $\Sigma(3000)$ resonance, respectively.
Experimental data are shown with the same symbols as in Fig.~\ref{FIG05}.}
\label{FIG08}
\end{figure*}

The effects of the resonance contributions are shown in Fig.~\ref{FIG06}.
The black dashed curve represents the Born-term contribution and is identical to the
black solid curve in Fig.~\ref{FIG02}.
The bump structures associated with the $\Sigma(2620)$ and $\Sigma(3000)$ resonances
are clearly visible.
In particular, the contribution from the latter is essential for describing the
available data.
Figures \ref{FIG06}(a) and \ref{FIG06}(b) show the results obtained with the $J^P$ =
$(1/2^+,\,3/2^+,\, 5/2^+)$ and $J^P$ = $(1/2^-,\, 3/2^-,\, 5/2^-)$ assignments,
respectively.

Figure~\ref{FIG07} displays the corresponding differential cross sections.
As already demonstrated in Fig.~\ref{FIG03}, the Born-term contribution alone
describes the experimental data only in the low-$-t'$ region.
Including the $\Sigma^*$ resonance contributions significantly improves the
agreement with the data at large $-t'$.
For the $J^P$ = $1/2^\pm$ assignments, the angular distributions are nearly flat,
whereas they become more structured as the spin increases.
The experimental data in the $\Sigma(3000)$ region are well reproduced for all
the spin-parity assignments considered, with particularly good agreement for
$J^P = 5/2^-$.

The effects of the individual $\Sigma^*$ contributions are more clearly illustrated
in Fig.~\ref{FIG08}, which shows the SDMEs at $P_{\mathrm{Lab}} = 4.25\,
\mathrm{GeV}$.
At this beam momentum, the resonance contribution is expected to arise predominantly
from the $\Sigma(3000)$.
The full results are obtained by adding the $\Sigma(3000)$ contribution to the
Born-term contribution shown in Fig.~\ref{FIG05}.
As discussed in Sec.~\ref{Sec:III-1}, the region $-t' \leqslant 1\, \mathrm{GeV}^2$
is well described by the $t$-channel processes.
In the region $-t' \geqslant 1\, \mathrm{GeV}^2$, some of the discrepancies between
the Born-term results and the experimental data are reduced by including the
$\Sigma(3000)$ contribution.

For $\rho_{11}$, the Born-term contribution, which is dominated by the $t$-channel
mechanism, already provides a satisfactory description over the entire region
$-t' \leqslant 2\, \mathrm{GeV}^2$.
For the $J^P$ = $3/2^\pm$ and $5/2^-$ assignments, the inclusion of the
$\Sigma(3000)$ contribution produces only a relatively small change in the full
result in both the helicity and GJ frames, thereby preserving the agreement achieved
by the Born term.
The $\rho_{11}$ results therefore favor the $J^P$ = $3/2^\pm$ and $5/2^-$
assignments.
For $\rho_{1-1}$, the Born-term result remains nearly constant at approximately 0.5
for $-t' \geqslant 0.5\, \mathrm{GeV}^2$.
After the $\Sigma(3000)$ contribution is included, the full result gradually
decreases with increasing $-t'$ in both the helicity and GJ frames for all the
considered spin-parity assignments.
This behavior is consistent with the experimental data.

For $\rho_{00}$, the available data are relatively limited~\cite{
Aguilar-Benitez:1972ngz,DeGroot:1974qw} and do not permit a definitive conclusion.
Only the $J^P = 1/2^\pm$ assignments reproduce the observed behavior of $\rho_{00}^H$
at $-t' \geqslant 1\, \mathrm{GeV}^2$, where the theoretical prediction increases
with increasing $-t'$ and is in agreement with the experimental data.
For Re\,$\rho_{10}$, all the considered $\Sigma(3000)$ assignments provide a good
description of the data in the region $-t' \leqslant 1.5\, \mathrm{GeV}^2$.
However, the data points at $-t' \simeq 1.8\, \mathrm{GeV}^2$~\cite{Rouge:1972xr}
are not reproduced by any of the considered assignments.

The SDMEs shown in Fig.~\ref{FIG04}, on the other hand, were measured at
$P_{\mathrm{Lab}} = 3.35\, \mathrm{GeV}$, where the contribution from the
$\Sigma(2620)$ resonance may be expected.
Nevertheless, the Born-term contribution alone provides a good description of the
experimental data over the entire $-t'$ region, indicating that the $\Sigma(2620)$
resonance plays a minor role in determining the SDMEs.
This is consistent with the smaller contribution of the $\Sigma(2620)$ to the total
cross section compared with that of the $\Sigma(3000)$, as shown in
Fig.~\ref{FIG06}.

\section{Summary and Conclusions}
\label{Sec:IV}

We investigated the $K^- p \to \phi \Sigma^0$ reaction within a hybrid Regge approach
based on effective Lagrangians.
The nonresonant background was described by the Reggeized $t$-channel $K$- and
$K^*$-meson exchanges, together with the $s$-channel ground-state $\Sigma$ and
$u$-channel nucleon exchanges.
We calculated the total and $t$-dependent differential cross sections as well as the
SDMEs in both the helicity and GJ frames, and compared the results with the available
experimental data.

The nonresonant reaction mechanism is governed predominantly by the $t$-channel
exchanges.
The $K^*$-Reggeon exchange provides the leading contribution to the
cross sections and accounts for their strongly forward-peaked behavior.
Its spin structure further indicates that the reaction is dominated by natural-parity
exchange.
Although smaller in magnitude, the K-Reggeon contribution remains essential for
describing the SDMEs, especially those sensitive to the helicity-zero and
unnatural-parity components.
The $s$-channel ground-state $\Sigma$ and $u$-channel nucleon contributions are found
to be negligible.
The combined $K$- and $K^*$-Reggeon exchanges provide a reasonable description of the
differential cross sections and of the SDMEs in the region $-t' \leqslant 1\,
\mathrm{GeV}^2$.

The nonresonant background alone, however, cannot reproduce the structures observed
in the total cross section for $3.0 \leqslant P_{\mathrm{Lab}} \leqslant 4.5\,
\mathrm{GeV}$, nor can it describe the differential cross sections at relatively
large $-t'$.
We therefore included the high-mass $\Sigma(2620)$ and $\Sigma(3000)$ resonances in
the $s$ channel and examined the possible spin-parity assignments $J^P=1/2^\pm$,
$3/2^\pm$, and $5/2^\pm$.
The $\Sigma(3000)$ contribution is particularly important for reproducing the
pronounced structure near $W \simeq 3.0\,\mathrm{GeV}$, whereas the effect of the
$\Sigma(2620)$ is comparatively modest.
The differential cross sections are reasonably well described for all the spin-parity
assignments considered, especially for the $J^P = 5/2^-$ assignment of the
$\Sigma(3000)$.
The SDMEs provide additional sensitivity to the resonance spin and parity.
In particular, the measured $\rho_{11}$ values at $P_{\mathrm{Lab}} = 4.25\,
\mathrm{GeV}$ tend to favor the $J^P = 3/2^\pm$ and $5/2^-$ assignments for the
$\Sigma(3000)$.
The $\rho_{1-1}$ data are also better reproduced when the $\Sigma(3000)$ contribution
is included.
On the other hand, the limited $\rho_{00}$ data at large $-t'$ appear to favor the
$J^P=1/2^\pm$ cases.
Taken together, these results suggest that the $5/2^-$ assignment is the most
plausible for the $\Sigma(3000)$, although the presently available data do not
allow a definitive determination.

Our results demonstrate that kaon-induced $\phi\Sigma^0$ production provides a
sensitive probe of high-mass $\Sigma^*$ resonances and their couplings to the
$\bar{K}N$ and $\phi\Sigma$ channels.
New measurements of the energy and angular dependences of the cross sections,
together with more precise SDMEs over a wider kinematic range, would be especially
valuable for clarifying the nature of the $\Sigma(2620)$ and $\Sigma(3000)$.
Such measurements could be carried out in future high-intensity kaon-beam experiments
at facilities such as J-PARC.

The present framework can also be extended to other vector- and axial-vector-meson
production reactions, such as $K^-p \to V\Lambda$ and $K^-p \to V\Sigma$, with
$V=\rho,\,\omega,\,a_1,\,f_1$, or $b_1$.
Such studies may help identify hyperon resonances that couple strongly to these
channels and thereby advance our understanding of hyperon spectroscopy.
Another interesting direction is to apply our framework to pion-induced reactions
$\pi^-p \to V n$, which may provide complementary information on the role of
intermediate $N^*$ resonances.
Investigations along these lines will be reported elsewhere.

\section*{Acknowledgments}

This work was supported by the National Research Foundation of Korea (NRF) grant
funded by the Korea government under Grants No. RS-2021-NR060129(SHK,
MKCh), RS-2026-25468435 (SHK), and RS-2025-16071941 (MKCh).



\end{document}